\documentclass[letterpaper,journal]{IEEEtran}
\usepackage{amsmath,amsfonts,amssymb}
\usepackage{algorithm}
\usepackage{algorithmicx}
\usepackage{algpseudocode}
\usepackage{array}
\usepackage[caption=false,font=normalsize,labelfont=sf,textfont=sf]{subfig}
\usepackage{textcomp}
\usepackage{stfloats}
\usepackage{url}
\usepackage{verbatim}
\usepackage{graphicx}
\usepackage{cite}
\usepackage{mathtools}
\usepackage{booktabs}
\usepackage{tabularx}
\usepackage{xcolor}
\newtheorem{theorem}{Theorem}
\long\def\old#1{}
\long\def\red#1{{\color{red}#1}}

\IEEEoverridecommandlockouts

\begin{document}

\title{Multi-Period 
Bilevel Coordination of Primary and Secondary Frequency Regulation for Grid-Forming Battery Energy Storage
%. in Islanded Microgrids
}

\author{Wenju Shang, Liangliang Hao, Jiangliang Jin, and Yunjian Xu%
	\thanks{W. Shang and L. Hao are with the School of Control Science and Engineering, Shandong University, Jinan, China.
	      J. Jin is with the School of Information and Intelligent Science, Donghua University, Shanghai, China.
		Y. Xu is with the Department of Mechanical Engineering and Automation, Chinese University of Hong Kong, Hong Kong SAR.
		Emails: llhao@sdu.edu.cn, wjshang@mail.sdu.edu.cn, jinjiangliang@dhu.edu.cn,    yjxu@mae.cuhk.edu.hk.
		\textit{(Corresponding author: Liangliang Hao)}}
}

\maketitle
\begin{abstract}
%We study the coordination of primary and secondary frequency regulation by grid-forming battery energy storage systems in islanded microgrids. 
%We formulate the coordination as a multi-period bilevel program, in which the upper level schedules secondary regulation setpoints and each lower level computes primary regulation decisions.
%Using the lower-level Karush--Kuhn--Tucker conditions, we recast the bilevel program as a single-level program with complementarity constraints that preserves its global optima.
%As standard constraint qualifications fail at every feasible point of such programs, we develop a barrier-coupled Scholtes--interior-point continuation method that ties the relaxation parameter to the interior-point barrier parameter and warm-starts successive stages. 
%For bounded iterates and multipliers with vanishing KKT
%residuals, we prove feasibility and C-stationarity
%of all accumulation points.
%On a frequency-critical islanded microgrids, numerical results show  that 
%the proposed method reduces
%solution time by 89.3\% and peak frequency deviation by 0.53\%
%relative to conventional Scholtes relaxation.

We study the coordination of primary and secondary frequency regulation by grid-forming battery energy storage systems in islanded microgrids. 
We formulate the coordination as a multi-period bilevel program, in which the upper level schedules secondary regulation setpoints and each lower level computes primary regulation decisions. 
Using the lower-level Karush–Kuhn–Tucker conditions, we obtain an exact single-level reformulation with complementarity constraints. 
To address the numerical difficulties introduced by these constraints, we develop a continuation method that couples the complementarity relaxation parameter with the interior-point barrier parameter and warm-starts successive stages. 
Under the stated boundedness assumptions and with vanishing subproblem residuals, we prove that every accumulation point is feasible and satisfies a first-order stationarity condition for the reformulated problem (C-stationarity). 
Numerical results for an islanded microgrid show that the proposed method reduces solution time by 89.3\% and peak frequency deviation by 0.53\% relative to conventional Scholtes relaxation.

\end{abstract}


\section{Introduction}
\label{sec:introduction}

%%%%%%%%%%%%%%%%%%%%%%%%%%%%%%%%%%%%%%%%%%%%%%%%%%%%
%%%%%%%%%%%%%%%%%%%%%%%%%%%%%%%%%%%%%%%%%%%%%%%%%%%%
\IEEEPARstart{A}{s} inverter-based renewable generation displaces synchronous generators, power systems lose inertia and frequency response capability~\cite{Wang2025VSG}. 
The loss of inertia is most acute in islanded microgrids, where local resources alone must balance load and renewable variations without support from a bulk grid. 
Since power imbalances can cause fast frequency excursions and high rates of change of frequency (RoCoF), frequency support must be both fast and sustained in microgrids~\cite{Heins2023}.

Grid-forming battery energy storage systems (BESSs) operated as virtual synchronous generators (VSGs) are well suited to frequency regulation. 
BESSs establish voltage and frequency references, provide virtual inertia and damping, and adjust their active power rapidly~\cite{Guan2022}. 
Primary frequency regulation (PFR) contains the initial frequency deviation, whereas secondary frequency regulation (SFR) restores the frequency over a slower timescale by updating power setpoints. 
The two layers are coupled: 
each SFR setpoint determines the power headroom available to PFR within its period, and the PFR actions in turn change the frequency, BESS power, and state of charge (SOC) from which the next period starts. 
%For heterogeneous BESSs, the quality of an SFR schedule depends on the constrained PFR response it induces.

{
Coordinating PFR and SFR in low-inertia microgrids poses three challenges.
First, fast PFR and slower SFR are coupled through the power constraints and SOC dynamics of heterogeneous BESSs.
Timescale separation alone does not capture this coupling \cite{10485493, FARAGALLA2025115467}.
Second, when the primary layer optimizes its response, SFR scheduling must anticipate this constrained optimum, and the terminal frequency, power, and SOC states also link successive periods.
Capturing these dependencies requires a multi-period hierarchical formulation \cite{11220941}.
Third, the upper-level objective depends implicitly on the optimal solutions of sequentially coupled lower-level problems. Changes in active  constraints in the lower level can render the upper-level objective nonsmooth \cite{MITZE2023110900, Dempe2025Bilevel}.
%Third, the scheduling objective depends implicitly on a chain of lower-level solutions, and active-set changes can make the response map nonsmooth \cite{MITZE2023110900, Dempe2025Bilevel}. 
%complicating efficient solution .
An efficient and reliable coordinated optimization method is needed to address these challenges.
}

\subsection{Related Works}
\label{sec:related_works}

Most existing studies focus on one regulation layer, with the other
represented by prescribed setpoints or control laws.
At the primary layer, an optimal coordinated control strategy is
proposed in~\cite{Zhu2019} to allocate regulation power among multiple
BESSs while incorporating SOC recovery.
A distributed online optimization method is developed
in~\cite{Zhao2020} to coordinate BESS frequency support.
In~\cite{SongHan2025PFR}, adaptive droop control is employed to improve
the response of large-scale storage.
A nonmonotonic droop design is proposed in~\cite{Alduaij2026} to expand
the feasible control region of inverter-based resources.
At the secondary layer, distributed averaging control is developed
in~\cite{SimpsonPorco2015} to restore frequency in islanded microgrids,
while robust load frequency control incorporating distributed BESSs
is investigated in~\cite{Oshnoei2020}.

Adaptive SFR incorporating primary dynamics is developed
in~\cite{Xue2024}, while deep reinforcement learning is employed
in~\cite{Ma2025} to jointly optimize PFR and SFR control parameters.
Predictive control is investigated for centralized transient
frequency regulation in~\cite{Heins2023} and hierarchical frequency
and tie-line power coordination in~\cite{MaBaiSaha2026}.
At the system level, inertia-dependent frequency response is
incorporated into stochastic unit commitment in~\cite{Teng2016FFR}.
Our focus is SFR scheduling that anticipates optimal PFR responses
under power and SOC constraints, with terminal states propagated
across successive scheduling periods.

Bilevel optimization provides a framework for SFR scheduling that
anticipates the optimal PFR response.
It is applied to storage participation in energy and regulation
markets in~\cite{Wang2022} and to distributed flexibility management
in~\cite{Fatemi2025}.
However, solving a bilevel program over multiple periods remains
computationally demanding.
The nested approaches discussed in~\cite{Sinha2018} require repeated
solutions of the lower-level problems.
A penalty alternating direction method is proposed
in~\cite{KleinertSchmidt2021}, which also involves repeated
subproblem solves.
Mixed-integer reformulations are discussed in~\cite{Kleinert2021},
but can introduce substantial combinatorial search.
For convex lower-level QPs with affine constraints, KKT-based
reformulations are considered in~\cite{Sinha2018,Liu2024}.
Under appropriate assumptions, the KKT conditions characterize
lower-level optimality and yield an exact single-level mathematical
program with complementarity constraints (MPCC).

The resulting MPCC, however, violates the Mangasarian--Fromovitz
constraint qualification at every feasible
point~\cite{Scholtes2001}.
Scholtes relaxation handles this degeneracy through
a sequence of relaxed nonlinear programs (NLPs).
Accumulation points of stationary solutions to these NLPs
are C-stationary under suitable MPCC constraint
qualifications~\cite{hoheisel2013theoretical}.
Repeated NLP solves can become costly as the number
of complementarity pairs grows.
%The effects of inexact solves on limiting stationarity are examined in~\cite{kanzow2015price}.
Barrier-driven interior-point methods are developed
in~\cite{RaghunathanBiegler2005,LeyfferLopezCalvaNocedal2006}
to coordinate relaxation tightening with barrier reduction.
With proportional coupling, however, a fixed parameter ratio
offers limited flexibility to tune early relaxation
and later tightening separately.
We design a saturating roll-off mapping to control
this transition for the multi-period MPCC.

%%%%%%%%%%%%%%%%%%%%%%%%%%%%%%%%%%%%%%%%%%%%%%%%%%%%%%%%%
\subsection{Summary of Contributions}
The main contributions are threefold.

First, we formulate PFR--SFR coordination among heterogeneous
grid-forming BESSs as a multi-period bilevel program.
The upper level schedules SFR setpoints, while each lower
level optimizes PFR commands under BESS dynamics and power
and SOC constraints.
The two levels retain distinct objectives for frequency
restoration and fast frequency support.
The optimal terminal frequency, power, and SOC states link
successive periods.
This structure enables SFR scheduling to anticipate both
the constrained optimal fast response and its effects on
subsequent operating conditions.

Second, 
we propose an exact single-level MPCC reformulation by
replacing the lower-level convex QPs with their complete
KKT conditions, including the boundary-condition multipliers.
The reformulation enforces inter-period state
continuity and enables simultaneous optimization across
the scheduling horizon.
We show that the proposed reformulation avoids nested lower-level QP solves during
upper-level optimization.
Under the stated equivalence assumptions, we establish
that it preserves the bilevel-feasible primal trajectories
and their objective values, and hence the global optimal
value of the original problem.

Third, we propose a barrier-coupled Scholtes--interior-point
continuation method for the resulting nonconvex MPCC.
A roll-off mapping coordinates relaxation tightening with
barrier reduction, while primal--dual warm starts and
slack reinitialization transfer solutions between stages.
For bounded primal iterates with vanishing barrier-KKT
residuals, we prove that every accumulation point is
MPCC-feasible and, under the stated multiplier bounds,
satisfies C-stationarity, a first-order stationarity condition
(cf. Theorem~\ref{thm:convergence}).
In the islanded microgrid case, the proposed method reduces
solution time by 89.3\% and peak frequency deviation by 0.53\%
relative to conventional Scholtes relaxation.

In what follows, Section~\ref{sec:model_formulation} presents the frequency regulation model and the bilevel problem. Section~\ref{sec:solution} develops the solution method, Section~\ref{sec:simulation} reports the simulation results, and Section~\ref{sec:conclusion} concludes the paper.

%%%%%%%%%%%%%%%%%%%%%%%%%%%%%%%%%%%%%%%%%%%%%%%%%%%%
%%%%%%%%%%%%%%%%%%%%%%%%%%%%%%%%%%%%%%%%%%%%%%%%%%%%	

\section{System Model and Bilevel Formulation}
\label{sec:model_formulation}

\subsection{Frequency Regulation Model}
\label{subsec:system_model}

\begin{figure}[!t]
	\centering
	\includegraphics[width=\columnwidth]{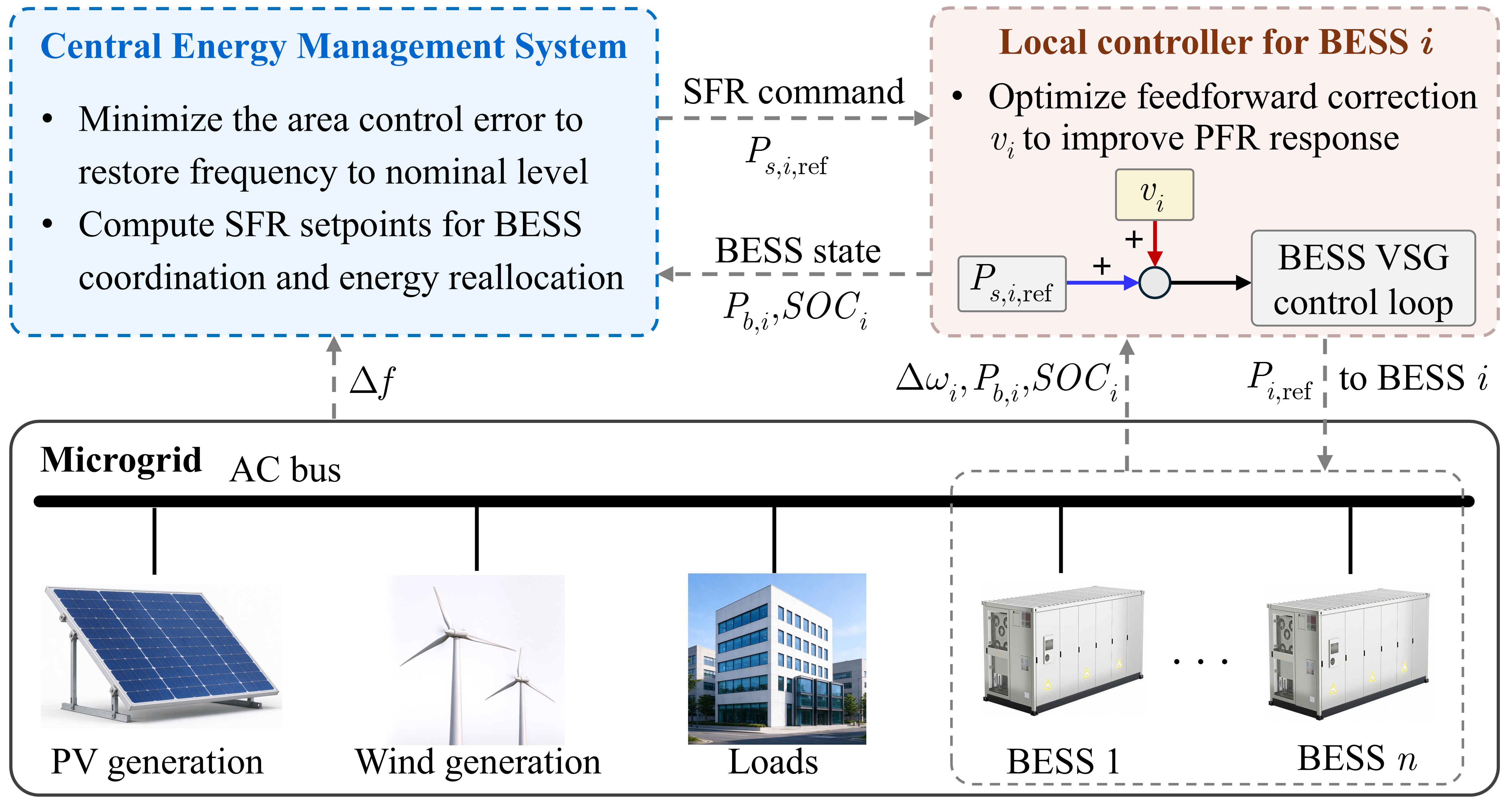}
	\caption{Coordinated primary and secondary frequency control strategy.}
	\label{fig:control_strategy}
\end{figure}

\red{\subsubsection{Control Architecture and Timescales}}
As shown in Fig.~\ref{fig:control_strategy}, 
we consider an islanded microgrid with $n$ heterogeneous grid-forming BESS units operated as VSGs, indexed by $i\in\mathcal{N}:=\{1,\dots,n\}$. The BESS units are the controllable regulation resources, whereas loads and renewable generation are nondispatchable. Under the coherent-frequency approximation, all units share the system frequency deviation $\Delta f_t$.
The central energy management system dispatches an SFR command to each BESS for frequency restoration, and the local controller superimposes the PFR command $v_{i,t}$ on it for fast frequency support. The VSG inertia and damping responses are generated internally, whereas $v_{i,t}$ is optimized by the PFR layer subject to power and SOC constraints.

\begin{figure}[!t]
	\centering
	\includegraphics[width=0.8\columnwidth]{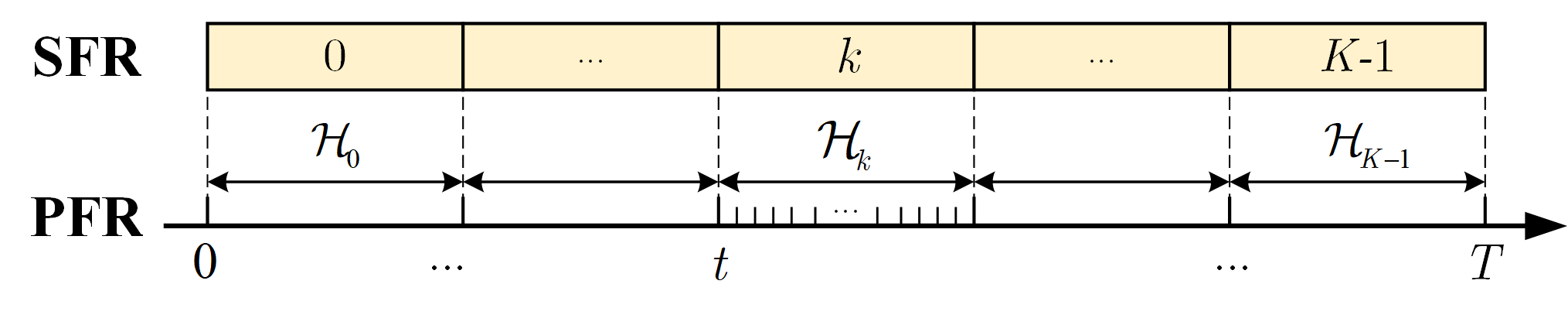}
	\caption{Operational timescales for SFR and PFR.}
	\label{fig:timescales}
\end{figure}

As shown in Fig.~\ref{fig:timescales}, the PFR layer operates with step size $\Delta t$ and time index $t$, whereas the SFR horizon is divided into $K$ periods of length $\Delta h$, indexed by $k$. We assume $\Delta h = \mathcal{T}\Delta t$, where $\mathcal{T}$ is the number of PFR steps per SFR period. Period $k$ thus covers the PFR index set
\begin{equation}
	\mathcal{H}_k \coloneqq \{\, t \mid k\mathcal{T} \leq t \leq (k+1)\mathcal{T} - 1 \,\}.
\end{equation}
Hence, any PFR index can be written as $t=k\mathcal{T}+\tau$ with $\tau\in\{0,\dots,\mathcal{T}-1\}$.

\red{\subsubsection{Grid-Forming BESS Model}}
Each BESS is connected to the alternating current (AC) microgrid through a voltage-source converter and an inductor-capacitor-inductor filter.
The active power reference of BESS $i$ combines the two commands:
\begin{equation}
	P_{i,\mathrm{ref},t}
	=
	P_{s,i,\mathrm{ref},t}
	+
	v_{i,t},
	\qquad
	\forall i\in\mathcal{N},\ \forall t,
	\label{eq:fr_law}
\end{equation}
where $P_{s,i,\mathrm{ref},t}$ is the SFR command. Since the SFR command is held constant within each period, i.e., $P_{s,i,\mathrm{ref},t}=P_{s,i,\mathrm{ref},k}$ for $t\in\mathcal{H}_k$, \eqref{eq:fr_law} becomes
\begin{equation}
	P_{i,\mathrm{ref},t}
	=
	P_{s,i,\mathrm{ref},k}
	+
	v_{i,t},
	\quad
	\forall i\in\mathcal{N},\ t\in\mathcal{H}_k.see
	\label{eq:pfr_law}
\end{equation}

\begin{figure}[!t]
	\centering
	\includegraphics[width=0.85\columnwidth]{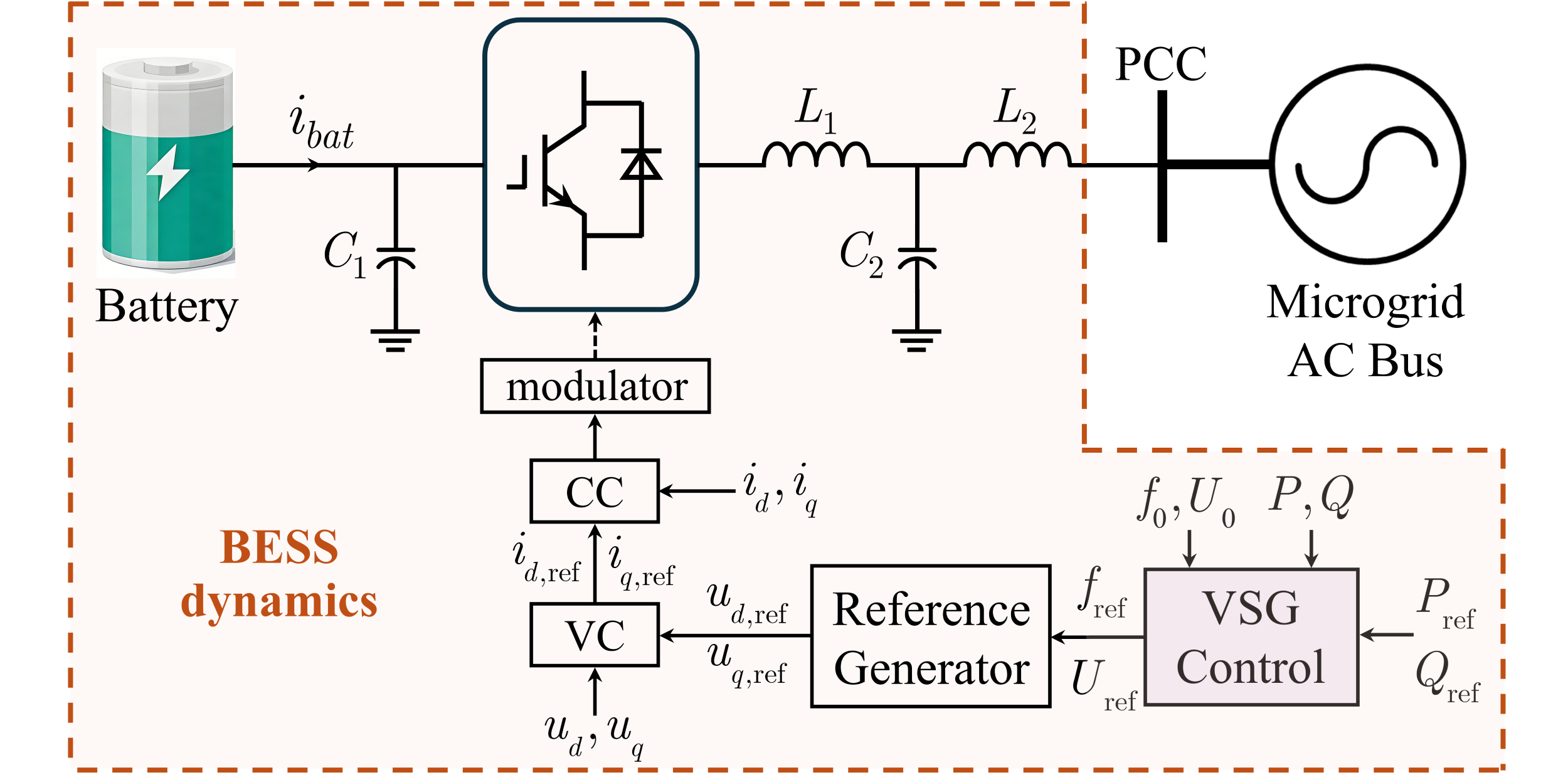}
	\caption{{Model of grid-forming BESS unit.}}
	\label{fig:bess_control}
\end{figure}

{Its control includes the power measurement, VSG, voltage control (VC), and current control (CC) loops (see Fig.~\ref{fig:bess_control}).}
Following reduced-order VSG active-power modeling \cite{Liang2024VSGReduced}, we approximate the reference-tracking dynamics by a second-order model with two cascaded first-order lags:
\begin{equation}
	\left[P_{\mathrm{a},i,t+1},P_{i,t+1}\right]^\top
	=
	A_i
	\left[P_{\mathrm{a},i,t},P_{i,t}\right]^\top
	+
	B_i P_{i,\mathrm{ref},t},
	\label{eq:bess_power_state}
\end{equation}
{where $P_{i,t}$ is the controllable power component and $P_{\mathrm{a},i,t}$ is an intermediate response state. The matrices $A_i\in\mathbb{R}^{2\times2}$ and $B_i\in\mathbb{R}^{2\times1}$ are obtained by exact zero-order-hold (ZOH) discretization of
	$T_{1,i}\dot P_{\mathrm{a},i} =P_{i,\mathrm{ref}}-P_{\mathrm{a},i}$ and $T_{2,i}\dot P_i=P_{\mathrm{a},i}-P_i$, i.e., $A_i=e^{A_{c,i}\Delta t}$ and $B_i=\int_0^{\Delta t}e^{A_{c,i}\tau}B_{c,i}\,d\tau$, with
	\[
	A_{c,i}=
	\begin{bmatrix}
		-1/T_{1,i} & 0\\
		1/T_{2,i} & -1/T_{2,i}
	\end{bmatrix},
	\quad
	B_{c,i}=
	\begin{bmatrix}
		1/T_{1,i}\\0
	\end{bmatrix}.
	\]
	%The effective response time constants $T_{1,i}$ and $T_{2,i}$ are listed in Table~\ref{tab:sim_parameters}.
	%The inherent VSG inertia and damping contributions to the terminal power are modeled separately in (4).
}

%The BESS dynamics reflect the coupled power measurement, VSG, voltage control (VC), and current control (CC) loops~\cite{Wang2025VSG}. To retain these aggregate converter control dynamics in the optimization formulation, the mapping from the external active power reference to the controllable power of BESS $i$ is represented by the following second order discrete-time model:
%\begin{equation}
%	\left[P_{\mathrm{a},i,t+1}, P_{i,t+1}\right]^\top
%	=
%	A_i
%	\left[P_{\mathrm{a},i,t}, P_{i,t}\right]^\top
%	+
%	B_i P_{i,\mathrm{ref},t},
%	\label{eq:bess_power_state}
%\end{equation}
%where $P_{i,\mathrm{ref},t}$ is the external active power reference, $P_{i,t}$ is the controllable output power, and $P_{\mathrm{a},i,t}$ is an intermediate response state. The matrices $A_i$ and $B_i$ are determined from the aggregate response parameters $(T_{1,i},T_{2,i})$ identified for BESS $i$.

The VSG controller of BESS $i$ also provides virtual inertia and damping, with the Hz-form coefficients
\begin{equation}
	J_i:=2\pi J_{\omega,i},
	\qquad
	\kappa_i:=2\pi D_{\omega,i}.
	\label{eq:Ji_kappa_definition}
\end{equation}
Here, $J_{\omega,i}$ and $D_{\omega,i}$ are the corresponding coefficients with respect to angular frequency. The inertial response depends on the RoCoF, defined as
\begin{equation}
	r_t = (\Delta f_{t+1}-\Delta f_t) / \Delta t.
	\label{eq:rocof_definition}
\end{equation}
The terminal power of BESS $i$ adds the damping and inertial responses to $P_{i,t}$:
\begin{equation}
	P_{b,i,t}
	=
	P_{i,t}
	-
	\kappa_i\Delta f_t
	-
	J_i r_t,
	\label{eq:terminal_power}
\end{equation}
where positive $P_{b,i,t}$ denotes discharging power injected into the microgrid. The SOC then evolves as
\begin{equation}
	\mathrm{SOC}_{i,t+1} = \mathrm{SOC}_{i,t} - \Delta t \frac{P_{b,i,t}}{3600E_{\mathrm{r},i}},
	\label{eq:dyn_soc}
\end{equation}
where $E_{\mathrm{r},i}$ is the rated energy capacity of BESS $i$.

\red{\subsubsection{Microgrid Frequency Dynamics}}
Aggregating the VSG coefficients of all units gives the equivalent inertia coefficient
\begin{equation}
	J_{\mathrm{sys}} = \sum_{i\in\mathcal{N}}J_i = 2\pi\sum_{i\in\mathcal{N}}J_{\omega,i},
	\label{eq:Jeq}
\end{equation}
and the equivalent damping coefficient
\begin{equation}
	D_{\mathrm{sys}} = D_L + \sum_{i\in\mathcal{N}}\kappa_i = D_L + 2\pi\sum_{i\in\mathcal{N}}D_{\omega,i},
	\label{eq:Dsys}
\end{equation}
where $D_L$ is the load-frequency damping coefficient, with $D_L=0$ for constant-power loads.

{
	The aggregate continuous-time frequency dynamics around the scheduled operating point are
	\begin{equation}
		J_{\mathrm{sys}}\dot{\Delta f}
		=\sum_{i\in\mathcal N}P_i-\Delta P_d
		-D_{\mathrm{sys}}\Delta f.
	\end{equation}
	With piecewise-constant net power over each sampling interval, exact ZOH discretization yields
}
\begin{equation}
	{
		\Delta f_{t+1}
		=
		a_f\Delta f_t
		+
		b_f
		\left(
		\sum_{i\in\mathcal N}P_{i,t}
		-
		\Delta P_{d,t}
		\right),
	}
	\label{eq:dyn_freq}
\end{equation}
where $a_f=e^{(-D_{\mathrm{sys}}\Delta t/J_{\mathrm{sys}})}$ and $b_f=(1-a_f)/D_{\mathrm{sys}}$ for
$D_{\mathrm{sys}}>0$ ($b_f=\Delta t/J_{\mathrm{sys}}$ when $D_{\mathrm{sys}}=0$).
Here, $\Delta P_{d,t}$ is the net active-power disturbance caused by load and renewable variations and incremental network losses.
%
%The microgrid frequency dynamics around the scheduled operating point are represented by the following aggregate model:
%\begin{equation}
%	\Delta f_{t+1}
%	=
%	\Delta f_t
%	+
%	\frac{\Delta t}{J_{\mathrm{sys}}}
%	\left(
%	\sum_{i\in\mathcal{N}}P_{i,t}
%	-
%	\Delta P_{d,t}
%	-
%	D_{\mathrm{sys}}\Delta f_t
%	\right),
%	\label{eq:dyn_freq}
%\end{equation}
%where $\Delta P_{d,t}$ denotes the net active power disturbance, including load variations, renewable generation variations, and incremental network losses referred to the aggregate power balance point. The equivalent inertia coefficient in the Hz-form model is

%Since $J_{\mathrm{sys}}$ and $D_{\mathrm{sys}}$ already include the VSG inertia and damping, the frequency dynamics~\eqref{eq:dyn_freq} use $P_{i,t}$, whereas the SOC dynamics~\eqref{eq:dyn_soc} use the terminal power $P_{b,i,t}$.

\subsection{Bilevel Optimization Formulation}
\label{subsec:bilevel_formulation}

%Based on the control strategy in Section~\ref{sec:frequency_control_model}, the coordinated primary and secondary frequency regulation problem naturally exhibits a hierarchical multi-timescale structure. A direct joint formulation over the complete horizon would tightly couple the slow-timescale SFR scheduling variables with all fast-timescale PFR response trajectories, leading to a large scale optimization problem. Therefore, we develop a bilevel framework. The upper level schedules SFR baselines, while the lower level determines the fast PFR response and returns the terminal state for inter-period coupling.

\subsubsection{Decision Variables}
The upper-level decision in SFR period $k$ is the setpoint vector
\begin{equation}
	\boldsymbol{u}_k := \{P_{s,i,\mathrm{ref},k}\}_{i=1}^{n}.
	\label{eq:u_def}
\end{equation}
The PFR state and the SFR boundary state are defined as
\begin{equation}
	\boldsymbol{x}_t
	:=
	\left[
	\Delta f_t,
	(P_{\mathrm{a},i,t}, P_{i,t}, \mathrm{SOC}_{i,t})_{i=1}^{n}
	\right]^\top
	\in \mathbb{R}^{3n+1},
	\label{eq:state_vector}
\end{equation}
and
\begin{equation}
	\boldsymbol{s}_k
	:=
	\left[
	s_{\Delta f}^{k},
	(s_{P_{\mathrm{a}},i}^{k}, s_{P,i}^{k}, s_{\mathrm{SOC},i}^{k})_{i=1}^{n}
	\right]^\top
	\in \mathbb{R}^{3n+1}.
	\label{eq:s_def}
\end{equation}
At each boundary instant $t=k\mathcal{T}$, the PFR state is linked to the boundary state by
\begin{equation}
	\boldsymbol{x}_{k\mathcal{T}}
	=
	\boldsymbol{s}_k,
	\qquad
	k=0,\dots,K-1.
	\label{eq:boundary_consistency}
\end{equation}
%Each lower-level problem is thus solved for given $\boldsymbol{u}_k$ and $\boldsymbol{s}_k$.

\subsubsection{Lower-Level PFR Response}
For given $(\boldsymbol{s}_k,\boldsymbol{u}_k)$, the lower level optimizes the PFR commands $v_{i,t}$, $t \in \mathcal{H}_k$, with the reference given by~\eqref{eq:pfr_law}. It minimizes the frequency deviation, RoCoF, and control effort:
\begin{subequations}\label{eq:lower_problem}
	\begin{align}
		\min_{{v_{i,t}, \boldsymbol x_t}} &\quad
		J_{p,k} := \omega_1\sum_{t \in \mathcal{H}_k}	\Big(\frac{\Delta f_t}{\Delta f_{\mathrm n}}\Big)^2\Delta t 	+ \omega_2 \sum_{t \in \mathcal{H}_k} \left( \frac{r_t}{r_{\mathrm n}} \right)^2 \Delta t \nonumber \\
		& + \rho_v \sum_{t \in \mathcal{H}_k} \sum_{i \in \mathcal{N}} v_{i,t}^2 \Delta t
		\label{eq:lower_obj}\\
		\text{s.t.}\quad
		& \eqref{eq:bess_power_state},\ \eqref{eq:terminal_power},\ \eqref{eq:dyn_soc},\ \eqref{eq:dyn_freq},\ \eqref{eq:pfr_law}, \quad \forall i \in \mathcal{N},\ t \in \mathcal{H}_k,
		\label{eq:lower_c1}\\
		& \underline{P}_{i,\mathrm{ref}} \le P_{i,\mathrm{ref},t} \le \overline{P}_{i,\mathrm{ref}}, \quad \forall i \in \mathcal{N},\ t \in \mathcal{H}_k,
		\label{eq:lower_c2}\\
		& \underline{\mathrm{SOC}}_i \le \mathrm{SOC}_{i,t} \le \overline{\mathrm{SOC}}_i, \nonumber \\
		&\forall i \in \mathcal{N},\ t\in\mathcal H_k^{\rm SOC},
		\label{eq:lower_c3}\\
		& \boldsymbol{x}_{k \mathcal{T}}= \boldsymbol{s}_{k},
		\label{eq:lower_c4}
	\end{align}
\end{subequations}
{where $\Delta f_n>0$ and $r_n>0$ normalize the frequency deviation and RoCoF, $\omega_1>0$ and $\omega_2>0$ are dimensionless weights, and $\rho_v>0$ weights the control effort.
The SOC limits are imposed on $\mathcal H_k^{\mathrm{SOC}} :=\{t\mid k\mathcal{T}+1\leq t\leq(k+1)\mathcal{T}\}$.
}

Problem~\eqref{eq:lower_problem} is a convex QP over a polyhedral feasible set. Since $\rho_v>0$ makes the reduced objective strongly convex in the PFR commands and the dynamics determine the states uniquely, its optimal trajectory is unique. We denote the resulting optimal terminal state by $\boldsymbol{x}^{*}_{(k+1)\mathcal{T}}$.

\subsubsection{Upper-Level SFR Scheduling}
To reserve headroom for the PFR response, the setpoints in~\eqref{eq:u_def} satisfy
\begin{equation}
	\underline{P}_{i,\mathrm{ref}}+\alpha_i \le P_{s,i,\mathrm{ref},k} \le \overline{P}_{i,\mathrm{ref}}-\alpha_i,
	\quad \forall i\in\mathcal{N},\ \forall k,
	\label{eq:sfr_headroom}
\end{equation}
where $\alpha_i\ge 0$ is the reserved headroom of BESS $i$. The optimal terminal state of each lower-level problem defines the next boundary state:
\begin{equation}
	\boldsymbol{s}_{k+1} := \boldsymbol{x}_{(k+1)\mathcal{T}}^{*}\!\left(\boldsymbol{s}_{k},\, \boldsymbol{u}_{k}\right),
	\quad k = 0,\dots, K-1.
	\label{eq:upper_link_map}
\end{equation}
The area control error (ACE) at the end of period $k$, i.e., at $t = (k+1)\mathcal{T}$, is defined as~\cite{Oshnoei2020}
\begin{equation}
	\mathrm{ACE}_{k+1} := \beta\,\Delta f_{(k+1)\mathcal{T}}^{*} + \Delta P_{\mathrm{tie},(k+1)\mathcal{T}}^{*},
	\label{eq:ace_def}
\end{equation}
where $\beta$ is the frequency bias factor and $\Delta P_{\mathrm{tie},(k+1)\mathcal{T}}^{*}$ is omitted under the single-area assumption. Hence, $\mathrm{ACE}_{k+1}$ depends on $(\boldsymbol{s}_k,\boldsymbol{u}_k)$ through~\eqref{eq:upper_link_map}.

Starting from
\begin{equation}
	\boldsymbol{s}_{0} = \boldsymbol{s}_{\mathrm{init}},
	\label{eq:upper_init}
\end{equation}
the upper-level problem over the $K$-period horizon is
\begin{subequations}\label{eq:upper_problem}
	\begin{align}
		\min_{\{\boldsymbol{u}_k\}_{k=0}^{K-1}}\quad
		& J_s := \sum_{k=0}^{K-1}\big(\mathrm{ACE}_{k+1}-\mathrm{ACE}_{\mathrm{ref}}\big)^2\,\Delta h
		\label{eq:upper_obj}\\
		\text{s.t.}\quad
		& \eqref{eq:sfr_headroom},\ \eqref{eq:upper_link_map},\ \eqref{eq:upper_init}. \label{eq:upper_c1}
	\end{align}
\end{subequations}
Let $J_{\mathrm {BL}}^\star$ denote the global optimal value of~\eqref{eq:upper_problem}. 
Changes in the lower-level active set can make the response
map nonsmooth, and the resulting upper-level problem
is generally nonconvex, which makes~\eqref{eq:upper_problem} difficult to solve. 
Section~\ref{sec:solution} replaces each lower-level problem with its KKT conditions to obtain a single-level MPCC.

\section{Method}
\label{sec:solution}

This section reformulates the bilevel problem as a single-level MPCC (Section~\ref{subsec:kkt_lower}), proposes the barrier-coupled continuation method (Section~\ref{subsec:improved_scholtes}), and analyzes its convergence (Section~\ref{subsec:convergence}).

\subsection{Single-Level MPCC Reformulation of the Bilevel Problem}
\label{subsec:kkt_lower}

Since each lower-level problem is a convex QP over a polyhedral
feasible set, its KKT conditions are necessary and sufficient for
global optimality. We therefore replace each lower-level problem with its KKT conditions.
%A key modeling detail is that the constraint $\boldsymbol{x}_{k\mathcal{T}}=\boldsymbol{s}_k$ links consecutive periods, and its multipliers enter the lower-level stationarity equations at $t=k\mathcal{T}$. The resulting KKT system therefore retains the inter-period coupling.

In period $k$, the lower-level primal variables are
$\Delta f_t$, $P_{\mathrm{a},i,t}$, $P_{i,t}$, and $\mathrm{SOC}_{i,t}$ for
$t\in\mathcal H_k\cup\{(k+1)\mathcal T\}$, and $v_{i,t}$ for $t\in\mathcal H_k$.
For compact notation, let $\boldsymbol p_{i,t} := \big[ P_{\mathrm{a},i,t}, P_{i,t} \big]^\top$, $C := \big[0,1\big]$, and $\Gamma_i := 3600E_{\mathrm{r},i}$, so that $P_{i,t}=C\boldsymbol p_{i,t}$. The frequency, BESS power, and SOC dynamics are then written as $h_{1,t}=0$, $\boldsymbol{h}_{2,i,t}=\boldsymbol{0}$, and $h_{3,i,t}=0$ for $i\in\mathcal N$ and $t\in\mathcal H_k$, with
\begin{equation}
		 {
			h_{1,t}:=
			\Delta f_{t+1}
			-
			a_f\Delta f_t
			-
			b_f
			\Big(
			\sum_{i\in\mathcal N}
			C\boldsymbol p_{i,t}
			-
			\Delta P_{d,t}
			\Big).
		}
	\label{eq:h_freq}
\end{equation}
%\begin{equation}
%	\begin{aligned}
%		h_{1,t} :&= \Delta f_{t+1}-\Delta f_{t} \\
%		&-\frac{\Delta t}{J_{\mathrm{sys}}}\Big(\sum_{i\in\mathcal{N}} C \boldsymbol{p}_{i,t}-\Delta P_{d,t}-D_{\mathrm{sys}}\Delta f_t\Big),
%	\end{aligned}
%	\label{eq:h_freq}
%\end{equation}
\begin{equation}
	\begin{aligned}
		\boldsymbol{h}_{2,i,t} := \boldsymbol{p}_{i,t+1} - A_i \boldsymbol{p}_{i,t} -B_i \Big(P_{s,i,\mathrm{ref},k}+v_{i,t}\Big),
	\end{aligned}
	\label{eq:h_bess}
\end{equation}
\begin{equation}
	\begin{aligned}
		h_{3,i,t} :&= \mathrm{SOC}_{i,t+1}-\mathrm{SOC}_{i,t} \\
		&+ \frac{\Delta t}{\Gamma_i}
		\big( C \boldsymbol{p}_{i,t} - \kappa_i \Delta f_t - J_i r_t \big).
	\end{aligned}
	\label{eq:h_soc}
\end{equation}

The inequality constraints~\eqref{eq:lower_c2} and~\eqref{eq:lower_c3} are written as $g_{j,i,t}\le0$, $j=1,\dots,4$, for $i\in\mathcal{N}$, with
\begin{align}
	g_{1,i,t} &:= \big(P_{s,i,\mathrm{ref},k} + v_{i,t} \big) - \overline{P}_{i,\mathrm{ref}},\quad \forall t \in \mathcal{H}_k, \label{eq:g_pref1} \\
	g_{2,i,t} &:= \underline{P}_{i,\mathrm{ref}} - \big(P_{s,i,\mathrm{ref},k} + v_{i,t} \big),\quad \forall t \in \mathcal{H}_k, \label{eq:g_pref2} \\
	g_{3,i,t} &:= \mathrm{SOC}_{i,t} - \overline{\mathrm{SOC}}_i,\quad \forall t\in\mathcal H_k^{\rm SOC}, \label{eq:g_soc1} \\
	g_{4,i,t} &:= \underline{\mathrm{SOC}}_i - \mathrm{SOC}_{i,t},\quad \forall t\in\mathcal H_k^{\rm SOC}. \label{eq:g_soc2}
\end{align}

We assign the multipliers $\lambda_{1,t}$,
$\boldsymbol{\lambda}_{2,i,t}\in\mathbb{R}^{2}$, and $\lambda_{3,i,t}$ to
\eqref{eq:h_freq}--\eqref{eq:h_soc}, respectively, and $\mu_{1,i,t}$, $\mu_{2,i,t}$ ($t\in\mathcal{H}_k$) and $\mu_{3,i,t}$, $\mu_{4,i,t}$ ($t\in\mathcal H_k^{\rm SOC}$) to \eqref{eq:g_pref1}--\eqref{eq:g_soc2}. The boundary condition~\eqref{eq:lower_c4} receives the multipliers
$\lambda_4^k$, $\boldsymbol{\lambda}_{5,i}^{k}\in\mathbb{R}^{2}$, and
$\lambda_{6,i}^k$ for
$\Delta f_{k \mathcal{T}}=s_{\Delta f}^k$,
$\boldsymbol{p}_{i,k \mathcal{T}}=\boldsymbol{s}_{p,i}^k$, and
$\mathrm{SOC}_{i,k \mathcal{T}}=s_{\mathrm{SOC},i}^k$, respectively, where $\boldsymbol{s}_{p,i}^{k}
:=\left[ s_{P_{\mathrm{a}},i}^{k}, s_{P,i}^{k} \right]^\top$.
{These boundary multipliers enter the stationarity conditions at $t=k\mathcal{T}$ and complete the KKT system of each period.}
The Lagrangian for period $k$ is
\begin{equation}
	\begin{aligned}
		\mathcal{L}_k
		&=
		J_{p,k}
		+ \sum_{t\in\mathcal{H}_k}\lambda_{1,t}\,h_{1,t}
		+ \sum_{t\in\mathcal{H}_k}\sum_{i\in\mathcal{N}} \boldsymbol{\lambda}_{2,i,t}^\top \boldsymbol{h}_{2,i,t} \\
		& + \sum_{t\in\mathcal{H}_k}\sum_{i\in\mathcal{N}}\lambda_{3,i,t}\,h_{3,i,t} + \lambda_{4}^{k}\big(\Delta f_{k\mathcal{T}}-s_{\Delta f}^{k}\big) \\
		& + \sum_{i\in\mathcal{N}} \left(\boldsymbol{\lambda}_{5,i}^{k}\right)^\top \big( \boldsymbol{p}_{i,k\mathcal{T}} - \boldsymbol{s}_{p,i}^{k}\big) \\
		& + \sum_{i\in\mathcal{N}}\lambda_{6,i}^{k}\big(\mathrm{SOC}_{i,k\mathcal{T}}-s_{\mathrm{SOC},i}^{k}\big) \\
		& + \sum_{t\in\mathcal{H}_k}\sum_{i\in\mathcal{N}}\mu_{1,i,t}\,g_{1,i,t}
		+ \sum_{t\in\mathcal{H}_k}\sum_{i\in\mathcal{N}}\mu_{2,i,t}\,g_{2,i,t} \\
		& + \sum_{t\in\mathcal H_k^{\rm SOC}}\sum_{i\in\mathcal{N}}\mu_{3,i,t}\,g_{3,i,t} + \sum_{t\in\mathcal H_k^{\rm SOC}}\sum_{i\in\mathcal{N}}\mu_{4,i,t}\,g_{4,i,t}.
	\end{aligned}
	\label{eq:lagrangian}
\end{equation}

The KKT conditions of~\eqref{eq:lower_problem} are as follows.

{(i) Primal feasibility:} All constraints in~\eqref{eq:lower_problem} hold.

{(ii) Dual feasibility ($\forall i\in\mathcal{N}$):}
\begin{align}
	\mu_{1,i,t}\ge 0,\ \mu_{2,i,t}\ge 0, \quad &\forall t\in\mathcal{H}_k, \label{eq:dual_pref}\\
	\mu_{3,i,t}\ge 0,\ \mu_{4,i,t}\ge 0, \quad &\forall t\in\mathcal H_k^{\mathrm{SOC}}. \label{eq:dual_soc}
\end{align}

{(iii) Complementary slackness ($\forall i\in\mathcal{N}$):}
\begin{equation}
	\mu_{j,i,t} g_{j,i,t} =0,
	\label{eq:comp_slackness}
\end{equation}
for $j\in\{1,2\}$, $t\in\mathcal H_k$, and for $j\in\{3,4\}$, $t\in\mathcal H_k^{\mathrm{SOC}}$.

{(iv) Stationarity:} $\nabla \mathcal{L}_k = 0$ with respect to all primal variables, where $t_0 := k\mathcal{T}$ and $t_f := (k+1)\mathcal{T}$.

With respect to $v_{i,t}$ ($t \in \mathcal{H}_k$):
\begin{equation}
	2\rho_v\,v_{i,t} \, \Delta t
	-B_i^\top \boldsymbol{\lambda}_{2,i,t}
	+\mu_{1,i,t}-\mu_{2,i,t}
	=0,\quad \forall i\in\mathcal{N}.
	\label{eq:stat_v}
\end{equation}

With respect to $\Delta f_t$:

\begin{equation}
{
	\begin{aligned}
		&2\omega_1\frac{\Delta f_{t_0}}
		{\Delta f_n^2} \Delta t
		-
		2\omega_2\frac{
			(\Delta f_{t_0+1}-\Delta f_{t_0})}
		{(\Delta t\,r_n)^2} \Delta t
		-a_f\lambda_{1,t_0}
		\\
		&+\lambda_4^k+
		\sum_{i\in\mathcal N}
		\lambda_{3,i,t_0}
		\frac{J_i-\Delta t\kappa_i}{\Gamma_i}
		=0
	\end{aligned}
}
\label{eq:stat_f0}
\end{equation}
%
%\begin{equation}
%	\begin{aligned}
%		&2\omega_1\frac{\Delta f_{t_0}}{\Delta f_{\mathrm n}^{2}}\Delta t
%		-2\omega_2\frac{\Delta f_{t_0+1}-\Delta f_{t_0}}{(\Delta t\,r_{\mathrm n})^{2}}\Delta t  \\
%		&+\lambda_{1,t_0}\Big(\frac{\Delta t D_{\mathrm{sys}}}{J_{\mathrm{sys}}}-1\Big) +\lambda_{4}^{k} + \sum_{i\in\mathcal{N}} \lambda_{3,i,t_0} \frac{J_i - \Delta t\kappa_i}{\Gamma_i}
%		=0,
%	\end{aligned}
%	\label{eq:stat_f0}
%\end{equation}
\begin{equation}
{
	\begin{aligned}
		&\frac{2\omega_1\Delta f_t\Delta t}
		{\Delta f_n^2}
		+
		\frac{2\omega_2
			(2\Delta f_t-\Delta f_{t+1}-\Delta f_{t-1})
			\Delta t}
		{(\Delta t\,r_n)^2}\\
		&+
		\lambda_{1,t-1}
		-a_f\lambda_{1,t}\\
		&+
		\sum_{i\in\mathcal N}
		\Big(
		\lambda_{3,i,t}
		\frac{J_i-\Delta t\kappa_i}{\Gamma_i}
		-
		\lambda_{3,i,t-1}
		\frac{J_i}{\Gamma_i}
		\Big)=0,\\
		&\quad
		t=t_0+1,\ldots,t_f-1.
	\end{aligned}
}
\label{eq:stat_fmid}
\end{equation}
%
%\begin{equation}
%	\begin{aligned}
%		&2\omega_1\frac{\Delta f_{t}}{\Delta f_{\mathrm n}^{2}}\Delta t
%		+2\omega_2\frac{2\Delta f_{t}-\Delta f_{t+1}-\Delta f_{t-1}}{(\Delta t\,r_{\mathrm n})^{2}}\Delta t \\
%		& +\lambda_{1,t-1} +
%		\lambda_{1,t}
%		\left(
%		\frac{\Delta t D_{\mathrm{sys}}}{J_{\mathrm{sys}}}
%		-1
%		\right) \\
%		& +
%		\sum_{i\in\mathcal N}
%		\left(
%		\lambda_{3,i,t}
%		\frac{J_i-\Delta t\kappa_i}{\Gamma_i}
%		-
%		\lambda_{3,i,t-1}
%		\frac{J_i}{\Gamma_i}
%		\right)
%		=0, \\
%		& \forall t=t_0+1,\dots,t_f-1,
%	\end{aligned}
%	\label{eq:stat_fmid}
%\end{equation}
\begin{equation}
	\begin{aligned}
		2\omega_2\frac{\Delta f_{t_f}-\Delta f_{t_f-1}}{(\Delta t\,r_{\mathrm n})^{2}}\Delta t
		+\lambda_{1,t_f-1}
		-\sum_{i\in\mathcal N} \lambda_{3,i,t_f-1} \frac{J_i}{\Gamma_i}
		=0.
	\end{aligned}
	\label{eq:stat_fT}
\end{equation}

With respect to $\boldsymbol{p}_{i,t}$ ($\forall i \in \mathcal{N}$):
\begin{equation}
{
	-A_i^\top\boldsymbol\lambda_{2,i,t_0}
	+
	C^\top
	\Big(
	-b_f\lambda_{1,t_0}
	+
	\lambda_{3,i,t_0}
	\frac{\Delta t}{\Gamma_i}
	\Big)
	+
	\boldsymbol\lambda_{5,i}^{k}
	=0.
}
\label{eq:stat_p0}
\end{equation}
%\begin{equation}
%	\begin{aligned}
%		&-A_i^\top \boldsymbol{\lambda}_{2,i,t_0} + C^\top \left( -\lambda_{1,t_0} \frac{\Delta t}{J_{\mathrm{sys}}} + \lambda_{3,i,t_0} \frac{\Delta t}{\Gamma_i} \right) \\
%		& + \boldsymbol{\lambda}_{5,i}^{k}=0,
%	\end{aligned}
%	\label{eq:stat_p0}
%\end{equation}
\begin{equation}
{
	\begin{aligned}
		&\boldsymbol\lambda_{2,i,t-1}
		-
		A_i^\top\boldsymbol\lambda_{2,i,t}+
		C^\top
		\Big(
		-b_f\lambda_{1,t}
		+
		\lambda_{3,i,t}
		\frac{\Delta t}{\Gamma_i}
		\Big)\\
		&=0,\quad
		t=t_0+1,\ldots,t_f-1.
	\end{aligned}
}
\label{eq:stat_pmid}
\end{equation}
%\begin{equation}
%	\begin{aligned}
%		&\boldsymbol{\lambda}_{2,i,t-1} - A_i^\top \boldsymbol{\lambda}_{2,i,t}
%		+ C^\top \big( -\lambda_{1,t} \frac{\Delta t}{J_{\mathrm{sys}}} \\
%		&+ \lambda_{3,i,t} \frac{\Delta t}{\Gamma_i} \big)
%		=0, \quad \forall t=t_0+1,\dots,t_f-1,
%	\end{aligned}
%	\label{eq:stat_pmid}
%\end{equation}
\begin{equation}
	\boldsymbol{\lambda}_{2,i,t_f-1}=0.
	\label{eq:stat_pT}
\end{equation}

With respect to $\mathrm{SOC}_{i,t}$ ($\forall i \in \mathcal{N}$):
\begin{equation}
	 -\lambda_{3,i,t_0}+\lambda_{6,i}^k =0,
	\label{eq:stat_soc0}
\end{equation}
\begin{equation}
	\lambda_{3,i,t-1}-\lambda_{3,i,t}+\mu_{3,i,t}-\mu_{4,i,t}=0,
	\quad \forall t=t_0+1,\dots,t_f-1,
	\label{eq:stat_socmid}
\end{equation}
\begin{equation}
	\lambda_{3,i,t_f-1}+\mu_{3,i,t_f}-\mu_{4,i,t_f}=0.
	\label{eq:stat_socT}
\end{equation}

%For each period $k$, we collect the lower-level variables over
%$\mathcal{H}_k\cup\{(k+1)\mathcal{T}\}$ as
{
For each period $k$, we collect the lower-level primal variables and associated KKT multipliers as follows:}
\begin{itemize}
	\item
	%1)  
	primal variables $\boldsymbol{\theta}_k := \{\Delta f_t,\boldsymbol{p}_{i,t},\mathrm{SOC}_{i,t}\}_{t\in\mathcal H_k\cup\{(k+1)\mathcal{T}\}} \cup \{v_{i,t}\}_{t\in\mathcal H_k}$,
	\item 
	%2) 
	equality multipliers $\boldsymbol{\lambda}_k := \{\lambda_{1,t},\, \boldsymbol{\lambda}_{2,i,t},\,\lambda_{3,i,t},\,\lambda_{4}^{k},\,\boldsymbol{\lambda}_{5,i}^{k},\,\lambda_{6,i}^{k}\}$,
	\item inequality multipliers $\boldsymbol{\mu}_k := \{\mu_{1,i,t},\,\mu_{2,i,t},\,\mu_{3,i,t},\,\mu_{4,i,t}\}$.
\end{itemize}
Let $\boldsymbol z$ collect all variables, namely the upper-level decisions $\{\boldsymbol{u}_k\}_{k=0}^{K-1}$, the boundary states, and the lower-level primal--dual variables. Replacing each lower-level problem~\eqref{eq:lower_problem} with its KKT conditions yields the single-level MPCC
\begin{subequations}\label{eq:mpcc}
	\begin{align}
		&\min_{\substack{
				\{\boldsymbol{u}_k\}_{k=0}^{K-1},\\
				\{\boldsymbol{s}_k\}_{k=0}^{K},\\
				\{\boldsymbol{\theta}_k,\boldsymbol{\lambda}_k,\boldsymbol{\mu}_k\}_{k=0}^{K-1}
		}}
		J_s(\boldsymbol z) := \sum_{k=0}^{K-1}\big(\mathrm{ACE}_{k+1}-\mathrm{ACE}_{\mathrm{ref}}\big)^2\,\Delta h
		\label{eq:mpcc_obj}\\
		& \text{s.t.}\quad
		\underline{P}_{i,\mathrm{ref}}+\alpha_i \le P_{s,i,\mathrm{ref},k} \le \overline{P}_{i,\mathrm{ref}}-\alpha_i, \; \forall i,k,
		\label{eq:mpcc_c1}\\
		& \qquad \boldsymbol{s}_0=\boldsymbol{s}_{\mathrm{init}},
		\label{eq:mpcc_c2}\\
		& \qquad \boldsymbol{s}_{k+1}=\boldsymbol{x}_{(k+1)\mathcal{T}}, \; \forall k,
		\label{eq:mpcc_c3}\\
		& \qquad \text{Eqs. } \eqref{eq:lower_c1}\text{--}\eqref{eq:lower_c4}, \eqref{eq:dual_pref}\text{--}\eqref{eq:stat_socT}
		\ \text{(KKT cond)},\; \forall k,
		\label{eq:mpcc_c4}
		\\
		& \qquad \mathrm{ACE}_{k+1} = \beta\, s_{\Delta f}^{k+1}, \; \forall k.
		\label{eq:mpcc_c5}
	\end{align}
\end{subequations}
{Constraints~\eqref{eq:lower_c4} and~\eqref{eq:mpcc_c3} enforce state continuity between consecutive SFR periods.}

Let $\mathcal F_{\mathrm{MPCC}}$ denote the feasible set of~\eqref{eq:mpcc}, with global optimal value
\begin{equation}
J_{\mathrm{MPCC}}^\star
:=\min_{\boldsymbol z\in\mathcal F_{\mathrm{MPCC}}}J_s(\boldsymbol z).
\label{eq:mpcc_optimal_value}
\end{equation}
Since the lower-level KKT conditions are necessary and sufficient,
\eqref{eq:mpcc} preserves the bilevel-feasible primal trajectories
and their objective values, so that $J_{\mathrm{MPCC}}^\star=J_{\mathrm{BL}}^\star$ (Appendix~\ref{app:convergence}).

\subsection{Barrier-Coupled Scholtes--Interior-Point Continuation Method}
\label{subsec:improved_scholtes}
 
{The complementarity constraints in Problem~\eqref{eq:mpcc} violate standard constraint qualifications at every feasible point and admit no strictly feasible point, so interior-point methods cannot be applied directly~\cite{Scholtes2001,hoheisel2013theoretical}. To relax the complementarity constraints, we first write~\eqref{eq:mpcc} in a generic form.} All equality constraints except the complementarity conditions are collected as
\begin{equation}
	\boldsymbol{H}(\boldsymbol{z})=\boldsymbol{0}.
\end{equation}
The complementarity conditions of the lower-level inequalities are written as
\begin{equation}
	\begin{aligned}
		&\pi_\ell(\boldsymbol z)\ge0,\quad
		w_\ell(\boldsymbol z)\ge0,\\
		&\pi_\ell(\boldsymbol z)w_\ell(\boldsymbol z)=0,
		\quad \ell=1,\ldots,N_c,
		\label{eq:generic_comp}
	\end{aligned}
\end{equation}
%\begin{equation}
%	0\le \pi_\ell(\boldsymbol{z}) \perp w_\ell(\boldsymbol{z})\ge 0,
%	\quad \ell=1,\dots, N_c,
%	\label{eq:generic_comp}
%\end{equation}
where $\pi_\ell$ is a lower-level inequality multiplier, $w_\ell=-g_\ell$ is the corresponding primal slack, and $N_c$ is the number of complementarity pairs.
 
Scholtes relaxation~\cite{Scholtes2001} replaces~\eqref{eq:generic_comp} by
\begin{equation}
	\pi_\ell(\boldsymbol z)\ge0, \quad
	w_\ell(\boldsymbol z)\ge0, \quad
	\pi_\ell(\boldsymbol z)w_\ell(\boldsymbol z)\le\xi,
	\label{eq:scholtes_relax}
\end{equation}
where $\xi>0$ is the relaxation parameter, and recovers~\eqref{eq:generic_comp} as $\xi\rightarrow0$.
For implementation, the relaxed product constraint is written as an equality with a slack $\sigma_\ell\ge0$, collected in $\boldsymbol \sigma$:
\begin{equation}
	\pi_\ell(\boldsymbol{z})w_\ell(\boldsymbol{z})+\sigma_\ell-\xi=0,
	\quad \sigma_\ell\ge0.
	\label{eq:scholtes_slack}
\end{equation}
 
The relaxed NLP is solved by a primal--dual interior-point method. For the analysis, the logarithmic-barrier subproblem at relaxation level $\xi$ and barrier parameter $\mu_{\mathrm B}>0$ is
\begin{equation}
	\begin{aligned}
		\min_{\boldsymbol z,\boldsymbol\sigma}\quad
		&
		J_s(\boldsymbol z)
		-
		\mu_{\mathrm B}
		\sum_{\ell=1}^{N_c}
		\left(
		\log \pi_\ell(\boldsymbol z)
		+
		\log w_\ell(\boldsymbol z)
		+
		\log \sigma_\ell
		\right) \\
		&-
		\mu_{\mathrm B}
		\sum_{m=1}^{N_b}
		\log q_m(\boldsymbol z)
		\\
		\mathrm{s.t.}\quad
		&
		\boldsymbol H(\boldsymbol z)=\boldsymbol0,
		\\
		&
		\pi_\ell(\boldsymbol z)w_\ell(\boldsymbol z)
		+
		\sigma_\ell
		-
		\xi
		=
		0,
		\qquad
		\ell=1,\dots,N_c,
	\end{aligned}
	\label{eq:barrier_relaxed_problem}
\end{equation}
where $q_m(\boldsymbol z)>0$, $m=1,\dots,N_b$, are the slacks of the inequality constraints outside the complementarity pairs, including the upper-level setpoint bounds. The equality multipliers of~\eqref{eq:barrier_relaxed_problem} are collected in $\boldsymbol y$, and the barrier terms induce the positive inequality multipliers $\mu_{\mathrm B}/\pi_\ell$, $\mu_{\mathrm B}/w_\ell$, $\mu_{\mathrm B}/\sigma_\ell$, and $\mu_{\mathrm B}/q_m$.
 
{We propose a barrier-coupled continuation method to solve
these subproblems efficiently.
It combines coupled parameter updates with primal--dual
warm starts and slack reinitialization.}

{\textit{1) Coupled update rule:}} The proposed method couples $\xi$ with $\mu_{\mathrm B}$ along a single continuation path.
%The coupled update synchronizes the reduction of complementarity violation and barrier perturbation, while the primal-dual transfer preserves local solution information between consecutive stages. This construction defines a single continuation path from a well-conditioned relaxed problem toward exact complementarity.
At continuation stage $r$, the barrier parameter is reduced superlinearly as
\begin{equation}
	\mu_{\mathrm B}^{r+1} = \alpha_\mu\left(\mu_{\mathrm B}^{r}\right)^{\beta_\mu},
	\label{eq:mu_update}
\end{equation}
where $0<\alpha_\mu<1$, $\beta_\mu>1$, and $0<\mu_{\mathrm B}^{0}<1$. The relaxation parameter then follows the roll-off mapping
\begin{equation}
	\xi^{r+1} = \psi(\mu_{\mathrm B}^{r+1}) = \xi_{\max} \frac{\left(\mu_{\mathrm B}^{r+1}\right)^{\gamma_\xi}} {\left(\mu_{\mathrm B}^{r+1}\right)^{\gamma_\xi}+c_\xi},
	\label{eq:xi_update}
\end{equation}
where $\xi_{\max}>0$ is the maximum relaxation level, $\gamma_\xi>0$ sets the steepness of the roll-off, and $c_\xi>0$ sets the roll-off scale, with $\psi(\mu_{\mathrm B})=\xi_{\max}/2$ at $\mu_{\mathrm B} = c_\xi^{1/\gamma_\xi}$.
While $\mu_{\mathrm B}\gg c_\xi^{1/\gamma_\xi}$, the mapping keeps $\xi\approx\xi_{\max}$, so the relaxation stays loose in the early stages. Afterwards, $\xi$ decreases approximately as $\xi_{\max}\mu_{\mathrm B}^{\gamma_\xi}/c_\xi$, so the relaxation is tightened together with the barrier rather than by a separate schedule. Since $\psi(0)=0$, exact complementarity is recovered as $\mu_{\mathrm B}\to0$.
 
{\textit{2) Slack reinitialization:}}
Reducing $\xi$ may cause the previous stage solution
to violate the tightened product bound.
Stage $r+1$ is warm-started with the primal variables
and equality multipliers from stage $r$.
The relaxation slacks are reinitialized as
\begin{equation}
\widehat{\sigma}_\ell^{\,r+1}\gets \max\!\big\{\xi^{r+1}- \pi_\ell(\boldsymbol z^r)w_\ell(\boldsymbol z^r),\ \epsilon_{\mathrm{int}}\big\},
\label{eq:sigma_reinit}
\end{equation}
where $\epsilon_{\mathrm{int}}>0$ is a small interior margin. The reinitialization keeps each stage initialization inside the barrier domain, even when the transferred iterate violates the tightened relaxation. The resulting equality residual is then removed by the infeasible-start primal--dual interior-point iterations.

% 
%Algorithm~\ref{alg:improved_scholtes} summarizes the finite-tolerance implementation. Hatted variables denote stage initializations, unhatted variables the corresponding stage solutions, and the interior push places all barrier slacks above $\epsilon_{\mathrm{int}}$.
 Algorithm~\ref{alg:improved_scholtes} summarizes the
finite-tolerance implementation.
Hatted variables denote stage initializations, and unhatted
variables denote stage solutions.
Lines~1--2 initialize the relaxation parameter and keep all
barrier slacks at least $\epsilon_{\mathrm{int}}$ from zero.
Each stage solves the barrier subproblem to accuracy
$\epsilon_r$ and evaluates the KKT and complementarity
residuals (lines~4--6).
The algorithm terminates when all stopping criteria are met
or the stage limit is reached (lines~7--10).
Otherwise, it updates the coupled parameters, transfers
the primal--dual solution, and reinitializes the relaxation
slacks for the next stage (lines~12--14).

\begin{algorithm}[t]
\caption{Barrier-coupled Scholtes--interior-point continuation method}
\label{alg:improved_scholtes}
	\begin{algorithmic}[1]
\Require $(\widehat{\boldsymbol z}^{\,0},\widehat{\boldsymbol y}^{\,0})$, $\mu_{\mathrm B}^{0}$, $\alpha_\mu$, $\beta_\mu$, $\xi_{\max}$, $\gamma_\xi$, $c_\xi$, $\epsilon_{\mathrm{int}}$, $\epsilon_{\mathrm{KKT}}$, $\epsilon_{\mathrm{comp}}$, $\xi_{\mathrm{tol}}$, $\{\epsilon_r\}$, and $R_{\max}$
		\Ensure Final iterate and termination status
		\Statex \textit{Initialization:}
		\State Set $\xi^0\gets\psi(\mu_{\mathrm B}^0)$ and apply an interior push so that $\pi_\ell(\widehat{\boldsymbol z}^{\,0}),w_\ell(\widehat{\boldsymbol z}^{\,0}),q_m(\widehat{\boldsymbol z}^{\,0})\ge\epsilon_{\mathrm{int}}$
		\State Set $\widehat{\sigma}_\ell^{\,0}\gets
		\max\!\big\{\xi^0-\pi_\ell(\widehat{\boldsymbol z}^{\,0})
		w_\ell(\widehat{\boldsymbol z}^{\,0}),\,
		\epsilon_{\mathrm{int}}\big\}$ for all $\ell=1,\ldots,N_c$
		\Statex \textit{Continuation:}
		\For{$r=0,\ldots,R_{\max}-1$}
		\State Solve \eqref{eq:barrier_relaxed_problem} at
		$(\mu_{\mathrm B}^r,\xi^r)$ to accuracy $\epsilon_r$
		\Statex \hspace{\algorithmicindent}from $(\widehat{\boldsymbol z}^{\,r},\widehat{\boldsymbol\sigma}^{\,r},
		\widehat{\boldsymbol y}^{\,r})$, and denote the solution by
		\Statex \hspace{\algorithmicindent}$(\boldsymbol z^r,\boldsymbol\sigma^r,\boldsymbol y^r)$
		\State Compute $\mathcal R_{\mathrm{KKT}}^r$ by~\eqref{eq:kkt_residual}
		\State $\mathcal R_{\mathrm{comp}}^r\gets\max_\ell\pi_\ell(\boldsymbol z^r)w_\ell(\boldsymbol z^r)$
		\If{$\xi^r\le\xi_{\mathrm{tol}}$ \textbf{and}
			$\mathcal R_{\mathrm{comp}}^r\le\epsilon_{\mathrm{comp}}$
			\textbf{and} $\mathcal R_{\mathrm{KKT}}^r\le\epsilon_{\mathrm{KKT}}$}
		\State \Return $(\boldsymbol z^r,\mathtt{converged})$
		\ElsIf{$r=R_{\max}-1$}
		\State \Return $(\boldsymbol z^r,\mathtt{stage\_limit})$
		\EndIf
		\State Update $(\mu_{\mathrm B}^{r+1},\xi^{r+1})$ by~\eqref{eq:mu_update}--\eqref{eq:xi_update}
		\State Set $(\widehat{\boldsymbol z}^{\,r+1},\widehat{\boldsymbol y}^{\,r+1})\gets(\boldsymbol z^r,\boldsymbol y^r)$
		\State Reinitialize $\widehat{\boldsymbol\sigma}^{\,r+1}$ by~\eqref{eq:sigma_reinit}
		\EndFor
	\end{algorithmic}
\end{algorithm}

\textbf{Remark 1:}
A standard implementation of Scholtes relaxation prescribes a decreasing sequence of relaxation parameters and solves each resulting relaxed NLP to full accuracy by an interior-point method with an independent barrier reduction. The proposed method departs from the standard scheme in two respects. First, the coupled update rule Eqs.~\eqref{eq:mu_update}--\eqref{eq:xi_update} replaces the prescribed relaxation sequence, so each stage solves a single barrier subproblem at $(\mu_{\mathrm B}^r,\xi^r)$ only to the accuracy $\epsilon_r$, and both parameters decrease along one continuation path. Second, the warm start with slack reinitialization Eq.~\eqref{eq:sigma_reinit} carries the primal variables and equality multipliers from one stage to the next while keeping each stage initialization strictly interior. The two features are designed to reduce the numbers of continuation stages and Newton iterations, and both reductions are observed in Section~\ref{subsec:computational_performance}.

\subsection{Convergence of the Proposed Method}
\label{subsec:convergence}

%For the limiting analysis, we remove the stopping tests and the stage limit of Algorithm~\ref{alg:improved_scholtes} and consider the resulting primal--dual stage sequence of~\eqref{eq:barrier_relaxed_problem}. In~\eqref{eq:mpcc}, $J_s$ is quadratic and $\boldsymbol H$, $\pi_\ell$, $w_\ell$, and $q_m$ are affine, so all of them are continuously differentiable.

{In this subsection, we analyze the convergence properties of the continuation
sequence as $r\to\infty$.}
Let $\mathcal F_{\mathrm{KKT}}$ denote the residual map of the primal-feasibility and stationarity equations of~\eqref{eq:barrier_relaxed_problem}, with arguments $(\boldsymbol z,\boldsymbol\sigma,\boldsymbol y,\mu_{\mathrm B},\xi)$. We define
\begin{equation}
	\mathcal R_{\mathrm{KKT}}^r
	:=
	\left\|
	\mathcal F_{\mathrm{KKT}}
	\left(
	\boldsymbol z^r,
	\boldsymbol\sigma^r,
	\boldsymbol y^r,
	\mu_{\mathrm B}^r,
	\xi^r
	\right)
	\right\|_\infty
	\label{eq:kkt_residual}
\end{equation}
as the residual at continuation stage $r$. Each stage iterate lies in the barrier domain, i.e., $\pi_\ell(\boldsymbol z^r)$, $w_\ell(\boldsymbol z^r)$, $\sigma_\ell^r$, and $q_m(\boldsymbol z^r)$ are strictly positive, and satisfies
\begin{equation}
	\mathcal R_{\mathrm{KKT}}^r\le\epsilon_r,
	\label{eq:inexact_conditions}
\end{equation}
where $\epsilon_r\to0$.

To characterize limiting stationarity, we define the effective multipliers
\begin{equation}
\begin{aligned}
a_\ell^r&:=
\frac{\mu_{\mathrm B}^r}{\pi_\ell(\boldsymbol z^r)}
-\frac{\mu_{\mathrm B}^r w_\ell(\boldsymbol z^r)}{\sigma_\ell^r},\\
b_\ell^r&:=
\frac{\mu_{\mathrm B}^r}{w_\ell(\boldsymbol z^r)}
-\frac{\mu_{\mathrm B}^r \pi_\ell(\boldsymbol z^r)}{\sigma_\ell^r}.
\end{aligned}
\label{eq:effective_multipliers}
\end{equation}
Under exact $\sigma_\ell$-stationarity, they are the coefficients of $-\nabla\pi_\ell$ and $-\nabla w_\ell$ after the multiplier of the relaxed equality is eliminated. A feasible point is C-stationary if it admits weak-stationarity multipliers with $a_\ell^\ast b_\ell^\ast\geq0$ for every biactive pair, i.e., every pair with $\pi_\ell(\boldsymbol z^\ast)=w_\ell(\boldsymbol z^\ast)=0$~\cite{hoheisel2013theoretical,kanzow2015price}.

{
\begin{theorem}
\label{thm:convergence}
Let $\{(\boldsymbol z^r,\boldsymbol\sigma^r,\boldsymbol y^r)\}$
be a sequence in the logarithmic-barrier domain satisfying
\eqref{eq:inexact_conditions} with $\epsilon_r\to0$.
Suppose the parameters follow
\eqref{eq:mu_update}--\eqref{eq:xi_update}
and $\{\boldsymbol z^r\}$ is bounded.
Then the following statements hold:

\textit{(i) Feasibility:}
Every accumulation point $\boldsymbol z^\ast$ is feasible
for the original MPCC~\eqref{eq:mpcc}:
\begin{equation}
\begin{aligned}
&\boldsymbol H(\boldsymbol z^\ast)=\boldsymbol0,\quad
q_m(\boldsymbol z^\ast)\geq0,\quad m=1,\ldots,N_b,\\
&\pi_\ell(\boldsymbol z^\ast)\geq0,\quad
w_\ell(\boldsymbol z^\ast)\geq0,\\
&\pi_\ell(\boldsymbol z^\ast)w_\ell(\boldsymbol z^\ast)=0,
\quad \ell=1,\ldots,N_c.
\end{aligned}
\label{eq:limit_comp}
\end{equation}

\textit{(ii) Objective bound:}
Consider any subsequence converging to $\boldsymbol z^\ast$
and retain the index $r$.
The objective values along this subsequence satisfy
\begin{equation}
	\lim_{r\to\infty}J_s(\boldsymbol z^r)
	=
	J_s(\boldsymbol z^\ast)
	\geq
	J_{\mathrm{MPCC}}^\star
	=
	J_{\mathrm{BL}}^\star.
	\label{eq:objective_limit}
\end{equation}

\textit{(iii) Stationarity:}
Suppose the equality multipliers associated with
$\boldsymbol H$, the effective multipliers
$\{a_\ell^r\}$ and $\{b_\ell^r\}$, and the ordinary-inequality
barrier multipliers
$\{\mu_{\mathrm B}^r/q_m(\boldsymbol z^r)\}$
remain bounded along the same subsequence.
Then $\boldsymbol z^\ast$ is C-stationary
for the original MPCC.
\end{theorem}
}

The proof is given in Appendix~\ref{app:convergence}.
Equation~\eqref{eq:limit_comp} establishes feasibility
for the original MPCC, including exact complementarity.
For the present model, this also preserves lower-level
PFR optimality and inter-period state continuity.
Equation~\eqref{eq:objective_limit} describes objective
convergence along each convergent subsequence.
%The limit need not equal the common global optimal value
of the two formulations.
The additional multiplier bounds establish C-stationarity,
a first-order stationarity condition.
These results support coupled parameter updates and
inexact stage solves with tolerances tending to zero.

\section{Simulation Results}
\label{sec:simulation}

This section evaluates the single-level MPCC reformulation
and the proposed continuation method

%This section validates the bilevel optimization framework and the proposed method in terms of computational efficiency and frequency regulation performance. 

\subsection{Simulation Setup and Benchmark Methods}
\label{subsec:simulation_setup}

\begin{figure}[!t]
	\centering
	\includegraphics[width=0.8\columnwidth]{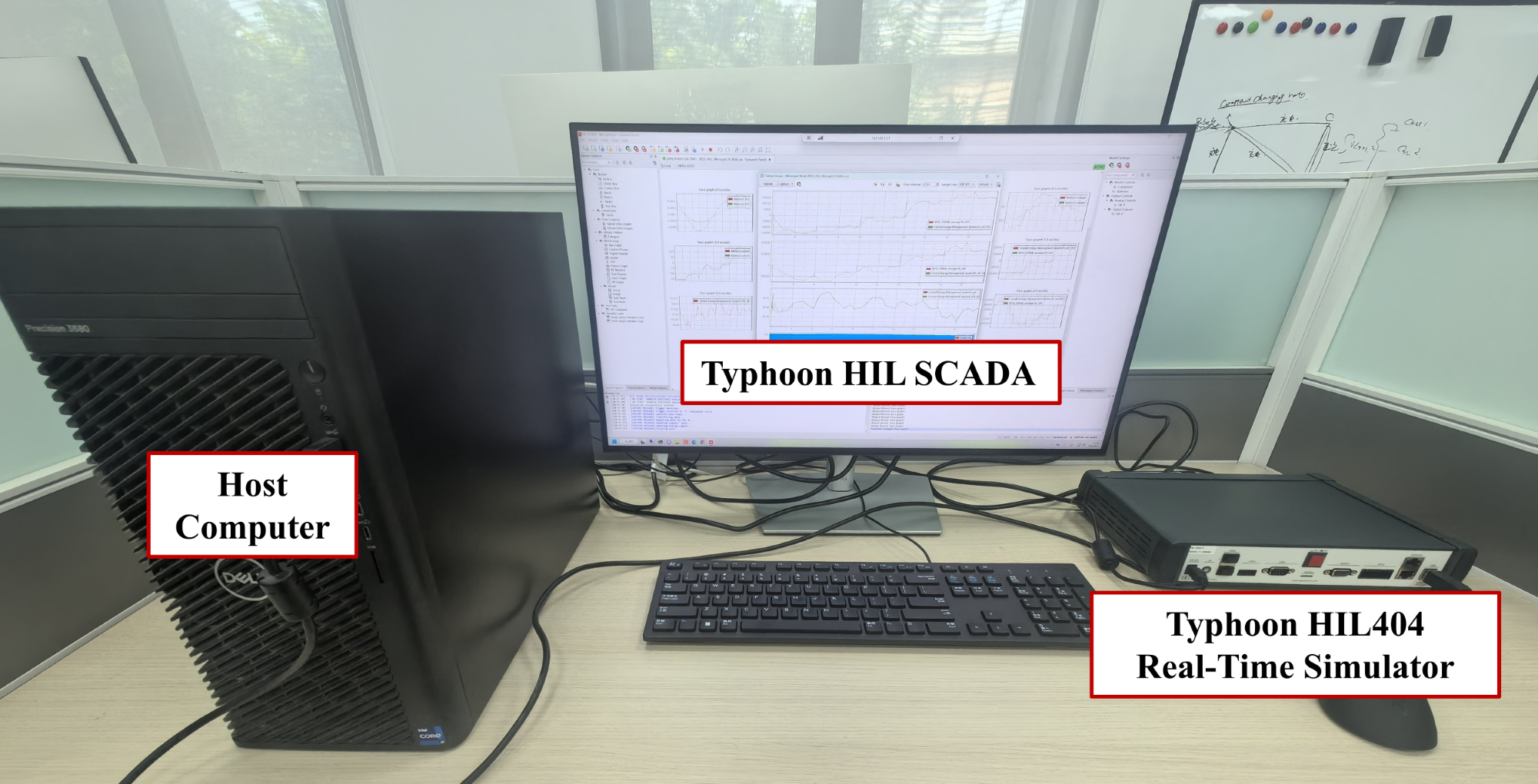}
	\caption{HIL test platform comprising the host computer, Typhoon HIL SCADA interface, and HIL 404 real-time simulator.}
	\label{fig:hil_platform}
\end{figure}

Fig.~\ref{fig:hil_platform} shows the HIL platform, comprising a host computer, the Typhoon HIL supervisory control and data acquisition (SCADA) interface, and a Typhoon HIL 404 real-time simulator. The SFR and PFR command trajectories are computed offline on the host computer and subsequently applied to the HIL model for real-time validation. The HIL 404 executes the microgrid model, while the SCADA interface records frequency, RoCoF, and BESS power responses.

\begin{figure}[!t]
	\centering
	\includegraphics[width=1.0\columnwidth]{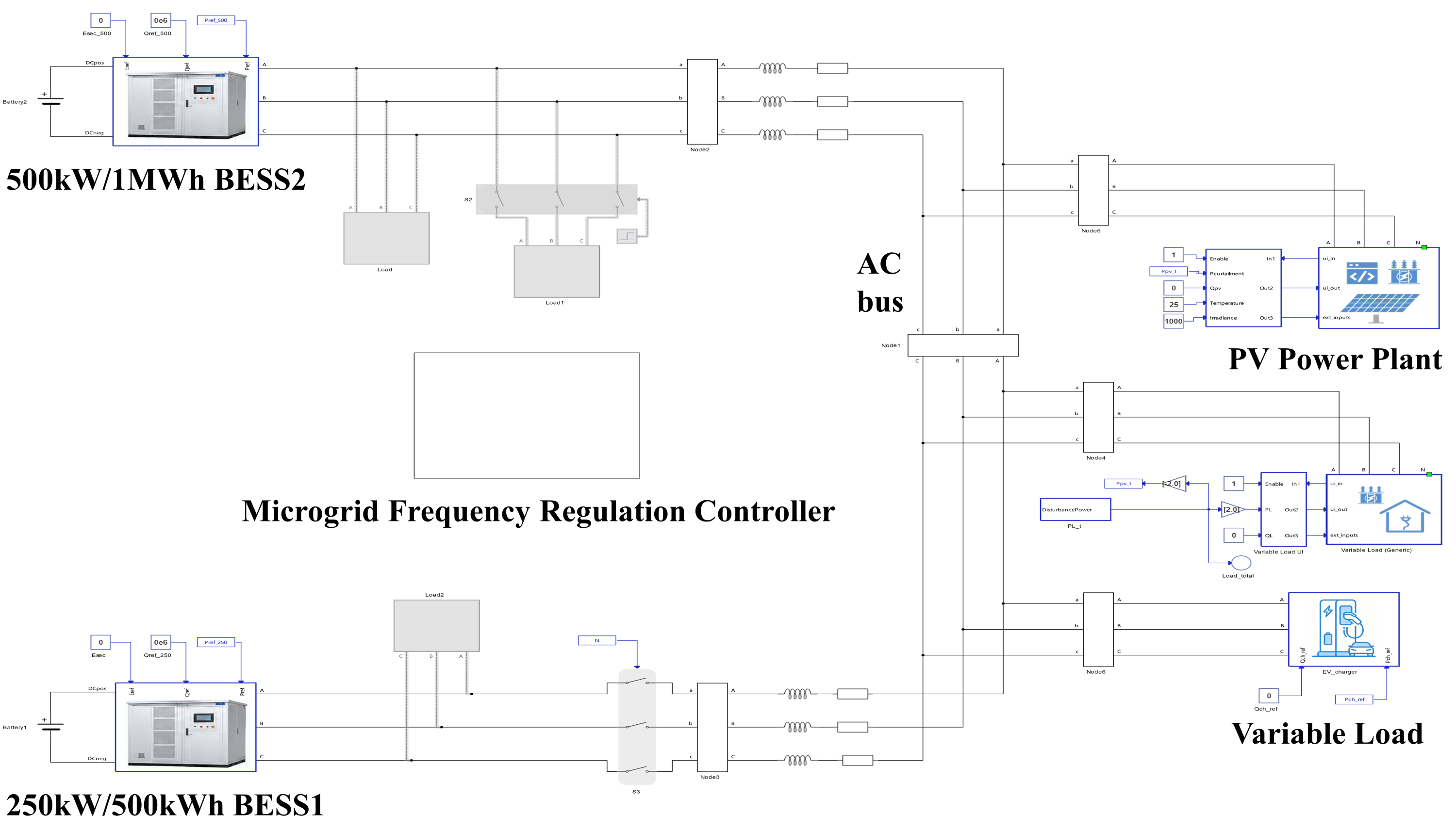}
		\caption{Detailed Typhoon HIL real-time model of the five-bus islanded microgrid with two heterogeneous grid-forming BESS units.}
	\label{fig:hil_model}
\end{figure}

The corresponding microgrid model is shown in Fig.~\ref{fig:hil_model}. The five-bus islanded system comprises 250kW/500kWh and 500kW/1MWh grid-forming BESS units, a photovoltaic power plant, and a variable load interconnected through the AC network. The controller applies the optimized SFR baselines and PFR commands to the two heterogeneous BESS units.
Its detailed converter, grid-forming control, and network dynamics provide a high-fidelity environment for evaluating the frequency regulation performance of the optimized commands.
%Its detailed converter, grid-forming control, and network dynamics provide a high-fidelity environment for validating the reduced-order model and coordinated decisions developed in Sections~\ref{sec:frequency_control_model}--\ref{sec:solution}.

Table~\ref{tab:sim_parameters} summarizes the main system and control parameters. The optimization horizon is 20s, and the HIL signals are recorded every 0.01s.
We compare the proposed method with grid search, mixed-integer quadratic programming (MIQP), alternating direction method of multipliers (ADMM), and conventional Scholtes relaxation. Grid search enumerates discretized SFR sequences and solves the lower-level QPs, MIQP encodes each complementarity pair with a binary variable and indicator constraints, and ADMM decomposes the regulation periods using locally linearized terminal-state mappings. Conventional Scholtes relaxation solves NLPs over a prescribed geometric parameter sequence.

\begin{table}[!t]
	\caption{Main Simulation Parameters}
	\label{tab:sim_parameters}
	\centering
	\footnotesize
	\setlength{\tabcolsep}{2.8pt}
	\renewcommand{\arraystretch}{1.10}
	\begin{tabularx}{\linewidth}{@{}lXlXlX@{}}
		\toprule
		Parameter & Value & Parameter & Value & Parameter & Value \\
		\midrule
		
		$\Delta t$
		& 0.1s
		& $\Delta h$
		& 4s
		& $\mathrm{ACE}_{\mathrm{ref}}$
		& 0MW \\
		
		$f_0$
		& 50Hz
		& $\beta$
		& 5MW/Hz
		& $D_L$
		& 0kW/Hz \\

		$\Delta f_{\mathrm n}$
		& 0.5Hz
		& $r_{\mathrm n}$
		& 1Hz/s
		& $T_{1,1}$
		& 0.204s \\

		$T_{2,1}$
		& 0.094s
		& $T_{1,2}$
		& 0.295s
		& $T_{2,2}$
		& 0.010s \\
		
		$\boldsymbol p_{i,0}$
		& $\boldsymbol 0$
		& $\mathrm{SOC}_{i,0}$
		& 0.5p.u.
		& $\overline{P}_{1,\mathrm{ref}}$
		& 250kW \\
		
		$\underline{P}_{1,\mathrm{ref}}$
		& -250kW
		& $\overline{P}_{2,\mathrm{ref}}$
		& 500kW
		& $\underline{P}_{2,\mathrm{ref}}$
		& -500kW \\
		
		$E_{\mathrm{r},1}$
		& 500kWh
		& $E_{\mathrm{r},2}$
		& 1000kWh
		& $\underline{\mathrm{SOC}}_i$
		& 0.2p.u. \\
		
		$\overline{\mathrm{SOC}}_i$
		& 0.9p.u.
		& $\alpha_1$
		& 100kW
		& $\alpha_2$
		& 150kW \\

		$J_{\omega,1}$
		& 15
		& $J_{\omega,2}$
		& 40
		& $D_{\omega,1}$
		& 250 \\
		
		$D_{\omega,2}$
		& 600
		& $\omega_1$
		& 1.0   
		&$\omega_2$
		& 10.0  \\
		
		$\rho_v$
		& 4.0 ${\mathrm{MW}}^{-2}$
		& $\alpha_\mu$
		& 0.35
		& $\beta_\mu$
		& 1.15 \\
		
		$\xi_{\max}$
		& 0.1
		& $\gamma_\xi$
		& 2.0
		& $c_\xi$
		& $10^{-5}$ \\
		\multicolumn{6}{@{}l@{}}{\scriptsize Units: $J_\omega$ in $\mathrm{kW\,s^2/rad}$, $D_\omega$ in $\mathrm{kW\,s/rad}$.} \\

		\bottomrule
	\end{tabularx}
\end{table}

All methods use the same disturbance trajectory, BESS parameters, operating constraints, and initial conditions and are implemented on a common computing platform with an Intel Core i9-14900K processor, 64GB of memory, and Windows 11 Pro.
Grid search evaluates five uniformly spaced candidates per BESS using 24 workers and solves the resulting QPs with OSQP at $10^{-6}$ absolute and relative tolerances. MIQP uses Gurobi indicator constraints, a 3600s limit, a $10^{-3}$ relative gap, and seed 1. ADMM uses at most 40 iterations, initial penalty 10, finite-difference step 0.03, and absolute/relative tolerances $10^{-5}$/$10^{-4}$. Conventional Scholtes relaxation uses 20 levels from $10^{-1}$ to $10^{-10}$, tolerances $10^{-7}$/$10^{-6}$, and at most 3000 iterations per subproblem. The proposed method uses~\eqref{eq:mu_update}--\eqref{eq:xi_update} with primal-dual warm starts.

\subsection{Convergence and Computational Performance}
\label{subsec:computational_performance}
	
\begin{figure}[!t]
	\centering
	\includegraphics[width=1.0\linewidth]{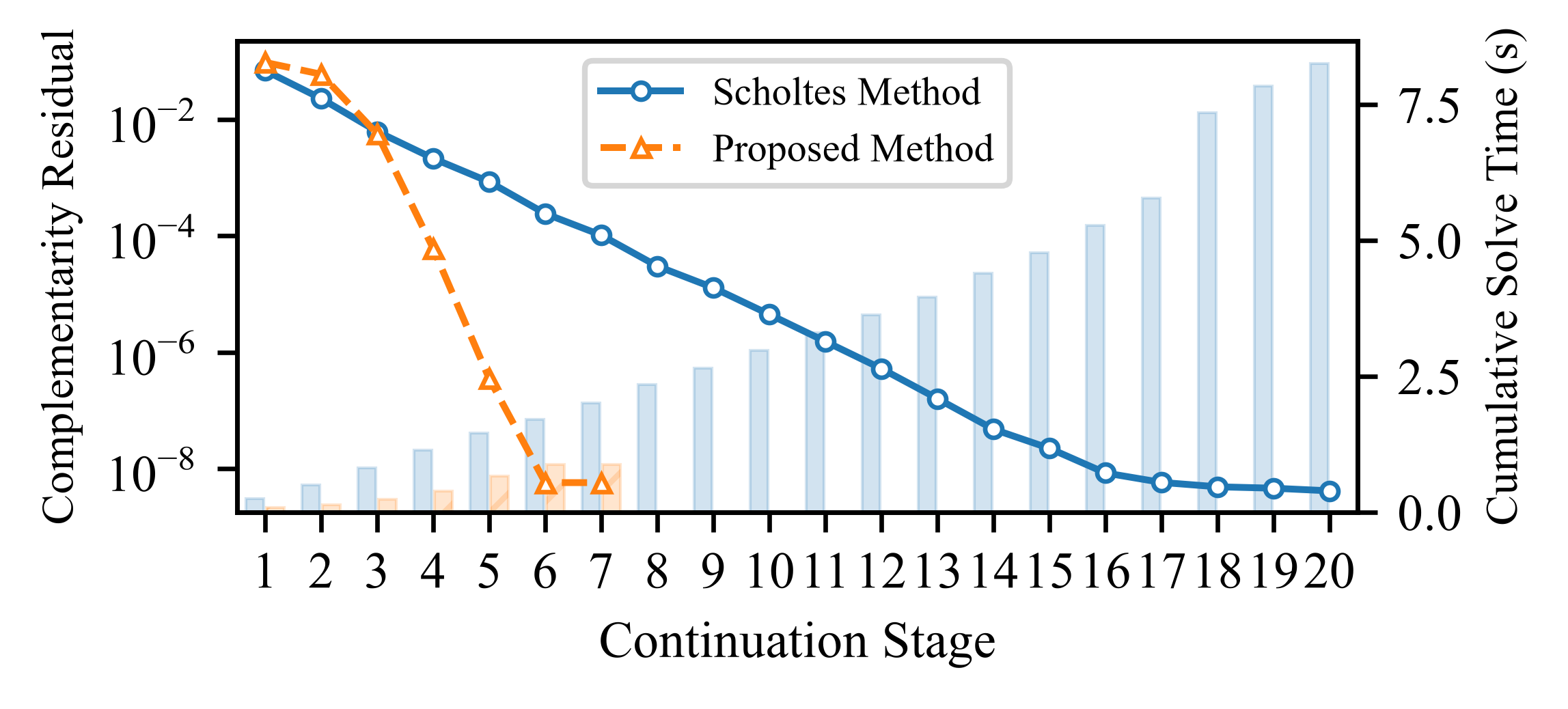}
\caption{Complementarity-residual convergence (lines, left axis) and cumulative solve time (bars, right axis) for conventional Scholtes relaxation and the proposed method.}
	\label{fig:convergence_cumulative_time}
\end{figure}

%We first compare the complementarity-residual behavior and cumulative solve time of conventional Scholtes relaxation and the proposed method. As shown in Fig.~\ref{fig:convergence_cumulative_time}, both methods reduce the maximum complementarity residual to the $10^{-8}$ level. The Scholtes method requires all 20 prescribed continuation stages, whereas the proposed method satisfies the termination criteria after seven stages, a reduction of 65\%. 
We first evaluate the convergence behavior and computational efficiency of the proposed method against conventional Scholtes relaxation. As shown in Fig.~\ref{fig:convergence_cumulative_time}, both methods reduce the maximum complementarity residual to the $10^{-8}$ level. The conventional method follows a prescribed sequence of 20 relaxation stages, whereas the proposed method satisfies the termination criteria within seven stages, reducing the number of stages by 65\%.
In addition, the cumulative-time bars show that most stages of the proposed method require less time than their conventional Scholtes counterparts, while the final tightening stage introduces only a small incremental cost. 
This behavior is consistent with the coupled path developed in Section~\ref{subsec:improved_scholtes}. The coupled $\mu_{\mathrm B}$-$\xi$ update avoids an unnecessarily dense sequence of intermediate relaxation levels, while the primal-dual continuation keeps consecutive iterates close to the relaxed central path and thereby reduces the required Newton corrections.
%This improvement is consistent with the proposed nonlinear roll-off mapping, which maintains a relatively relaxed complementarity bound during the early stages and accelerates relaxation tightening as the barrier parameter decreases. Combined with primal-dual warm starts and relaxation-slack reinitialization, the coordinated continuation strategy facilitates successive NLP solves while avoiding unnecessary intermediate relaxation stages.

	Table~\ref{tab:performance_comparison} reports solve times of 0.8852s and 8.2571s for the proposed and conventional Scholtes methods, respectively, corresponding to an 89.3\% reduction. The reductions relative to grid search, MIQP, and ADMM are 99.96\%, 98.1\%, and 96.6\%, respectively. The dominant computational burdens clarify this ordering. Let $N_i$ be the number of SFR candidates for BESS $i$ in one period and $N_{\mathrm{GS}}:=\prod_{i=1}^{n}N_i$ be the joint candidate count. If $T_{\mathrm{QP,ll}}$ is the cost of one lower-level QP evaluation, grid search has worst-case work $\mathcal{O}\!\left(N_{\mathrm{GS}}^K T_{\mathrm{QP,ll}}\right)$, which grows combinatorially with the candidate count, number of BESS units, and number of SFR periods.
	For the ADMM implementation, let $R_{\mathrm{ADMM}}$ denote the iteration count. Each iteration requires $K$ nominal lower-level evaluations and approximately $Kn$ additional evaluations for finite-difference linearization of the terminal-state mappings. Its dominant burden is $\mathcal{O}\!\left(R_{\mathrm{ADMM}}K(n+1)T_{\mathrm{QP,ll}}\right)$. MIQP introduces one binary variable for each of the $\mathcal N_c=4KnT$ complementarity pairs, and its worst-case branch-and-bound burden is exponential in $N_c$. In this case, $K=5$, $n=2$, and $\mathcal{T}=40$, resulting in 1600 binary variables before presolve.

	%Table~\ref{tab:performance_comparison} reports $0.8852\mathrm{s}$ and $8.2571\mathrm{s}$ for the proposed and conventional Scholtes methods, an 89.3\% reduction. The reductions relative to grid search, MIQP, and ADMM are 99.96\%, 98.1\%, and 96.6\%, respectively. The dominant computational burdens clarify this ordering. Let $N_i$ be the candidate count for BESS $i$, $N_{\mathrm{GS}}:=\prod_iN_i$, and $T_{\mathrm{QP,ll}}$ the cost of one lower-level QP evaluation. Grid search requires $\mathcal{O}(N_{\mathrm{GS}}^K T_{\mathrm{QP,ll}})$ work. With $R_{\mathrm{ADMM}}$ iterations, ADMM requires $\mathcal{O}(R_{\mathrm{ADMM}}K(n+1)T_{\mathrm{QP,ll}})$ because of the finite-difference terminal-state linearization. MIQP introduces one binary variable for each of the $N_c=2Kn(2\mathcal{T}+1)$ complementarity pairs, giving 1600 binary variables before presolve for $K=5$, $n=2$, and $\mathcal{T}=40$, with a worst-case branch-and-bound burden exponential in $N_c$.
	
	For conventional Scholtes relaxation and the proposed method, let $R_m$ be the number of continuation stages and $I_m^{(r)}$ the number of primal-dual Newton iterations at stage $r$, where $m\in\{\mathrm{Sch},\mathrm{prop}\}$. If $T_{\mathrm{KKT}}$ denotes the dominant cost of one Newton step, then the total computational burden is
	$\mathcal{O}\!\left(\sum_{r=1}^{R_m} I_m^{(r)}T_{\mathrm{KKT}}\right)$.
	The parameter coupling and primal-dual warm starts reduce continuation stages and Newton corrections, giving $R_{\mathrm{prop}} < R_{\mathrm{Sch}}$ and $\sum_{r=1}^{R_{\mathrm{prop}}}I_{\mathrm{prop}}^{(r)} < \sum_{r=1}^{R_{\mathrm{Sch}}}I_{\mathrm{Sch}}^{(r)}$. In this case, the empirical totals are $\sum_{r=1}^{R_{\mathrm{Sch}}}I_{\mathrm{Sch}}^{(r)} = 203$ and $\sum_{r=1}^{R_{\mathrm{prop}}}I_{\mathrm{prop}}^{(r)} = 27$, corresponding to an 86.7\% reduction in cumulative Newton iterations.
	Overall, the proposed method reduces the continuation stages and Newton effort, avoids the repeated finite-difference QP evaluations required by ADMM, and introduces neither the binary variables nor branch-and-bound search required by the MIQP formulation. These structural differences explain the runtime ordering in Table~\ref{tab:performance_comparison}.

	%For conventional Scholtes relaxation and the proposed method, let $R_m$ be the number of continuation stages and $I_m^{(r)}$ the number of primal-dual Newton iterations at stage $r$, where $m\in\{\mathrm{Sch},\mathrm{prop}\}$. If $T_{\mathrm{KKT}}$ denotes the dominant cost of one Newton step, then the total computational burden is $\mathcal{O}\!\left(\sum_{r=1}^{R_m} I_m^{(r)}T_{\mathrm{KKT}}\right)$. The parameter coupling and primal-dual warm starts reduce continuation stages and Newton corrections, giving $R_{\mathrm{prop}} < R_{\mathrm{Sch}}$ and $\sum_{r=1}^{R_{\mathrm{prop}}}I_{\mathrm{prop}}^{(r)} < \sum_{r=1}^{R_{\mathrm{Sch}}}I_{\mathrm{Sch}}^{(r)}$. In this case, the empirical totals are $\sum_{r=1}^{R_{\mathrm{Sch}}}I_{\mathrm{Sch}}^{(r)} = 203$ and $\sum_{r=1}^{R_{\mathrm{prop}}}I_{\mathrm{prop}}^{(r)} = 27$, corresponding to an 86.7\% reduction in cumulative Newton iterations. The lower continuation effort and the absence of finite-difference QP evaluations or binary variables explain the runtime ordering in Table~\ref{tab:performance_comparison}.

	\begin{table}[!t]
		\caption{Comparison of Computational and Dynamic Performance}
		\label{tab:performance_comparison}
		\centering
		\footnotesize
		\setlength{\tabcolsep}{3.0pt}
		\begin{tabularx}{\linewidth}{@{}l*{3}{>{\centering\arraybackslash}X}@{}}
			\toprule
		Method & Time (s) & $\max_t|\Delta f_t|$ (Hz) & $\max_t|r_t|$ (Hz/s) \\
			\midrule
		VSG-only & -- & 0.07567 & 0.4870 \\
		Grid search & 2234.7 & 0.01737 & 0.4687 \\
		MIQP & 47.732 & 0.01720 & 0.5102 \\
		ADMM & 26.133 & 0.01708 & 0.4775 \\
		Scholtes & 8.2571 & 0.01703 & 0.4691 \\
		\textbf{Proposed method} & \textbf{0.8852} & \textbf{0.01694} & \textbf{0.4607} \\
			\bottomrule
		\end{tabularx}
	\end{table}
	
	\subsection{Frequency Regulation Performance}
	
\begin{figure}[!t]
		\centering
		\includegraphics[width=1.0\linewidth]{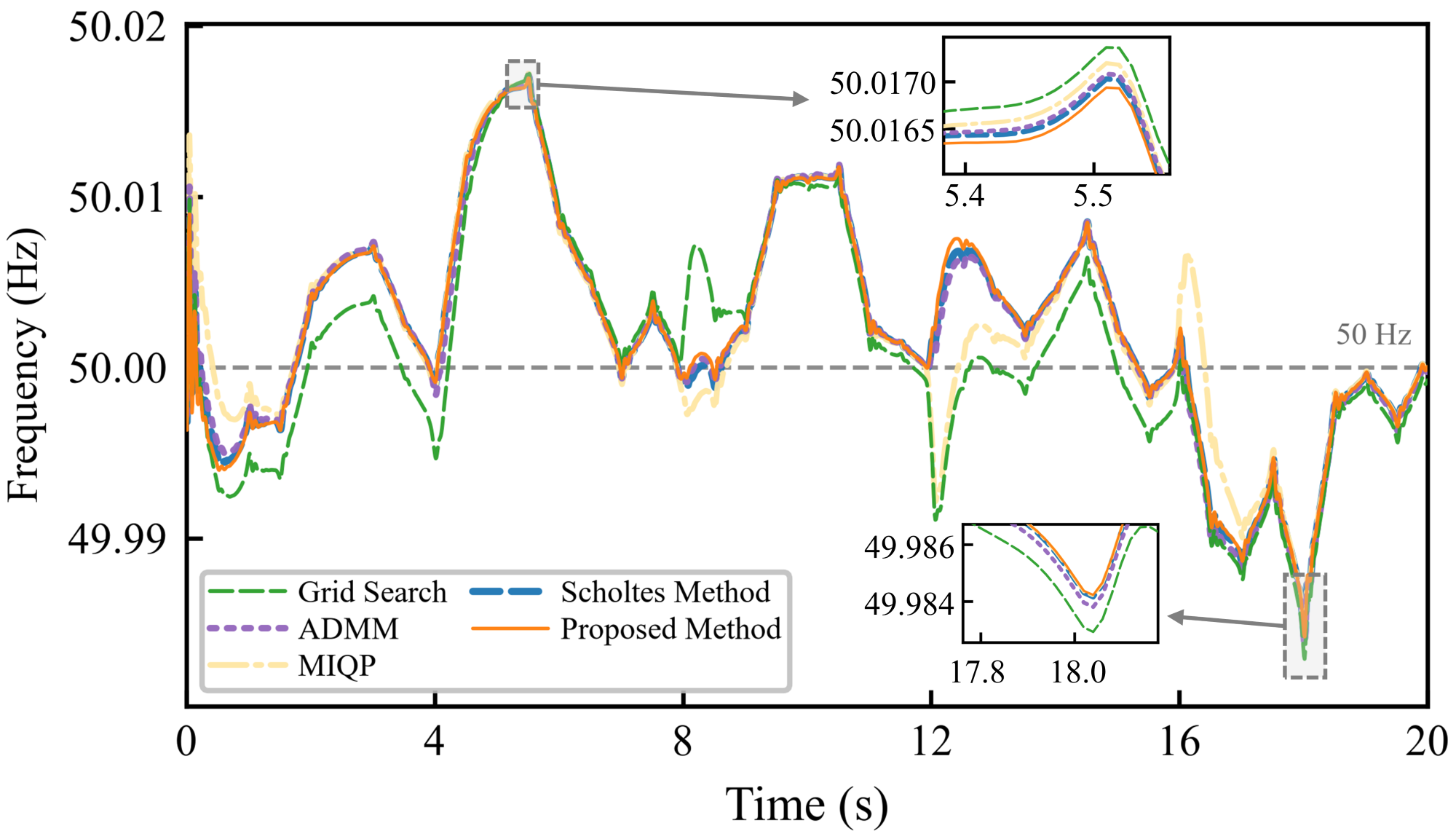}
		\caption{HIL frequency responses over the 20s disturbance trajectory.}
		\label{fig:frequency_response}
	\end{figure}

	We next evaluate the optimized commands in the HIL model. Fig.~\ref{fig:frequency_response} compares the resulting frequency responses. The unoptimized trajectory is omitted to preserve the visible differences among the optimized cases, while its peak metrics are retained in Table~\ref{tab:performance_comparison}.
	
	Grid search exhibits a slightly larger frequency excursion than the other optimized cases, consistent with the discretization imposed by its finite upper-level search grid.
	The enlarged regions near 5.5s and 18.0s show that the proposed method yields the smallest deviations around the local peak and nadir.
	As summarized in Table~\ref{tab:performance_comparison}, the proposed method achieves the lowest maximum absolute frequency deviation of 0.01694Hz. This represents a reduction of 77.6\% relative to the VSG-only operation and reductions ranging from 0.53\% to 2.48\% relative to the four optimization-based benchmarks.

\begin{figure}[!t]
		\centering
		\includegraphics[width=1.0\linewidth]{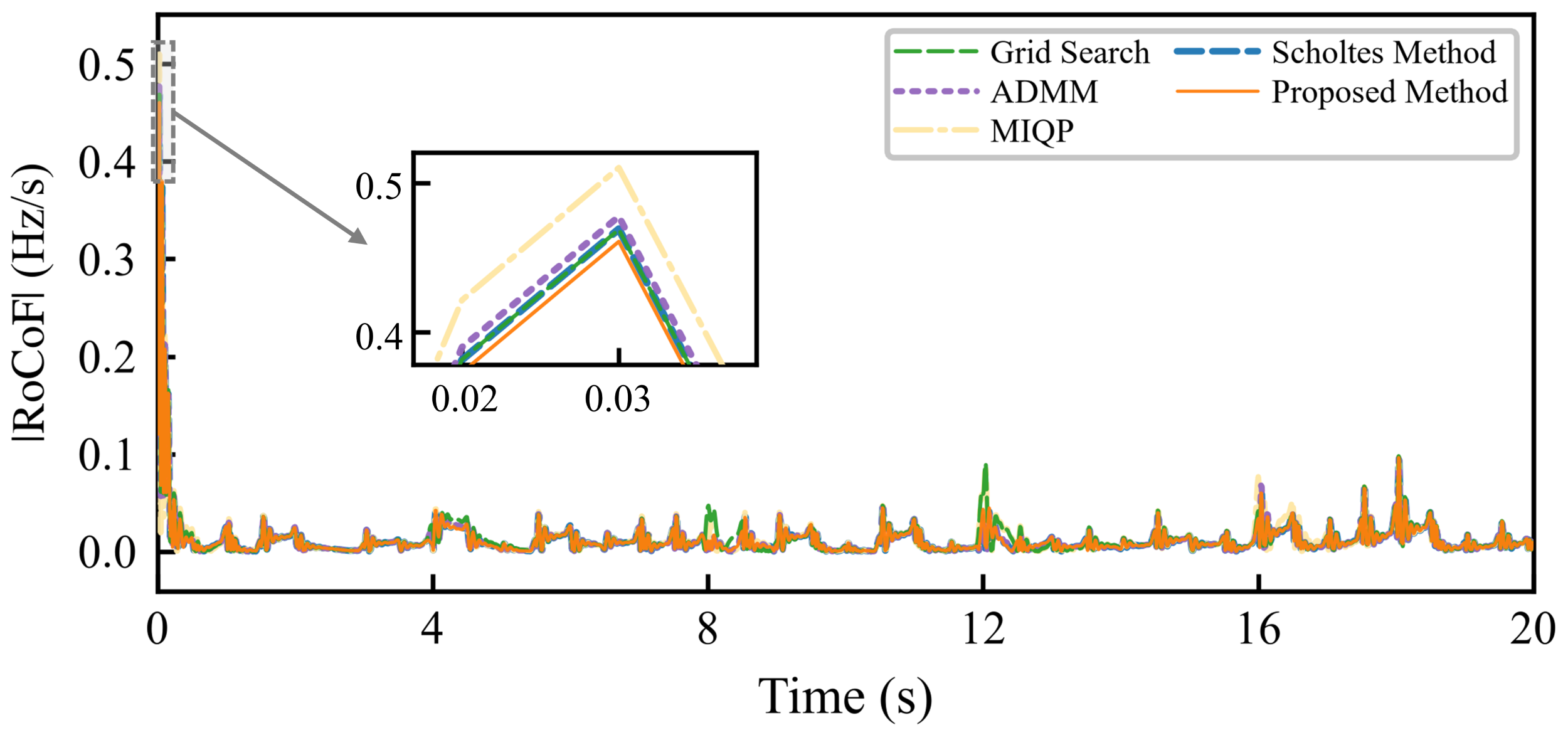}
		\caption{Absolute RoCoF responses over the 20s HIL disturbance trajectory.}
		\label{fig:rocof_response}
	\end{figure}

%	The absolute RoCoF responses in Fig.~\ref{fig:rocof_response} further show that the proposed method achieves the smallest maximum absolute RoCoF, 0.4607Hz/s. This value is 5.40\% lower than that of the unoptimized case and 1.71\%--9.70\% lower than those of the optimization-based benchmarks. These results demonstrate improved dynamic frequency regulation performance with substantially less computation.
	The absolute RoCoF responses in Fig.~\ref{fig:rocof_response} further show that the proposed method achieves the lowest reported peak value of 0.4607~Hz/s, a 5.40\% reduction relative to VSG-only operation and 1.71\%--9.70\% lower than the other four optimization-based benchmarks. Together, these results demonstrate effective frequency regulation by the optimized commands and comparable or slightly improved dynamic performance across the optimization-based methods. Importantly, the proposed method achieves this performance with a substantially lower computational burden, as demonstrated in Section~\ref{subsec:computational_performance}.
	
	\subsection{Multi-Timescale BESS Coordination}
	\label{subsec:bess_coordination}
	
\begin{figure}[!t]
		\centering
		\includegraphics[width=1.0\linewidth]{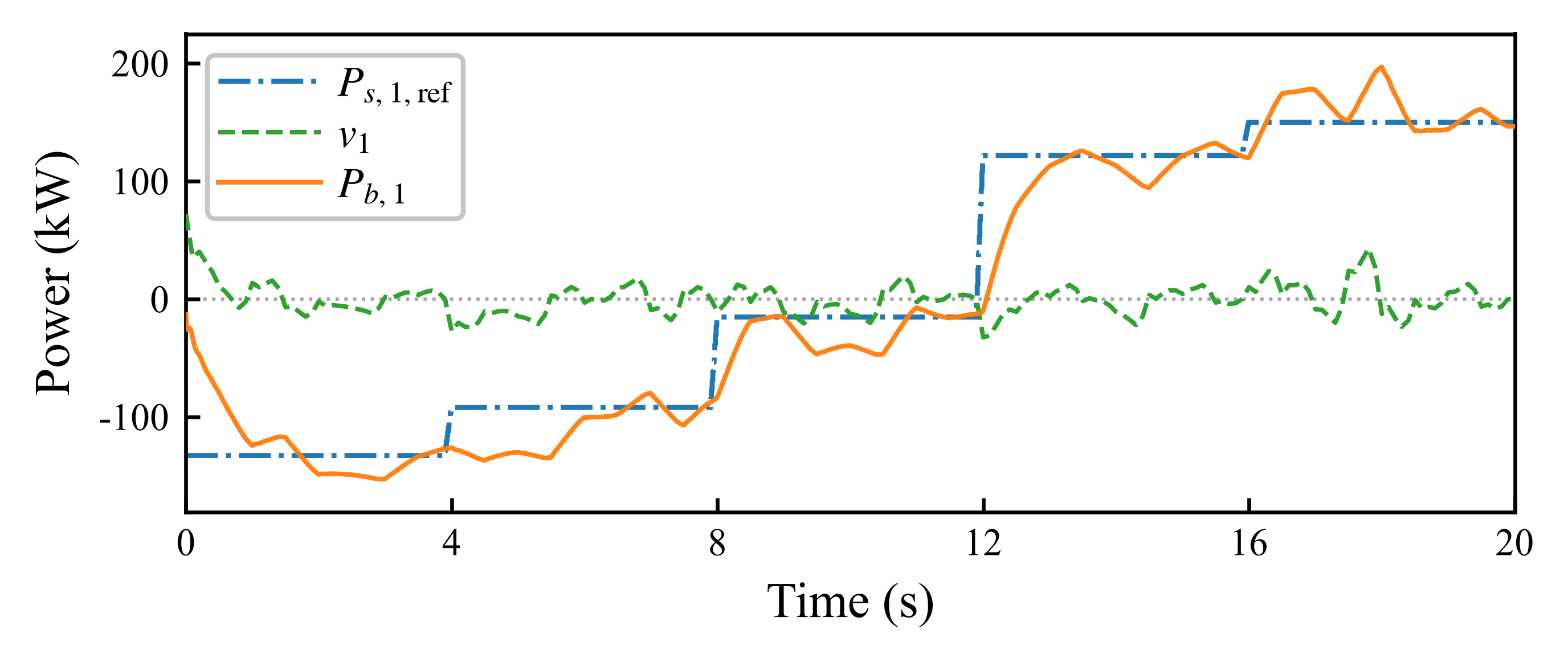}
		\caption{SFR baseline, PFR command, and terminal power of BESS $1$.}
		\label{fig:bess1_power}
	\end{figure}

\begin{figure}[!t]
		\centering
		\includegraphics[width=1.0\linewidth]{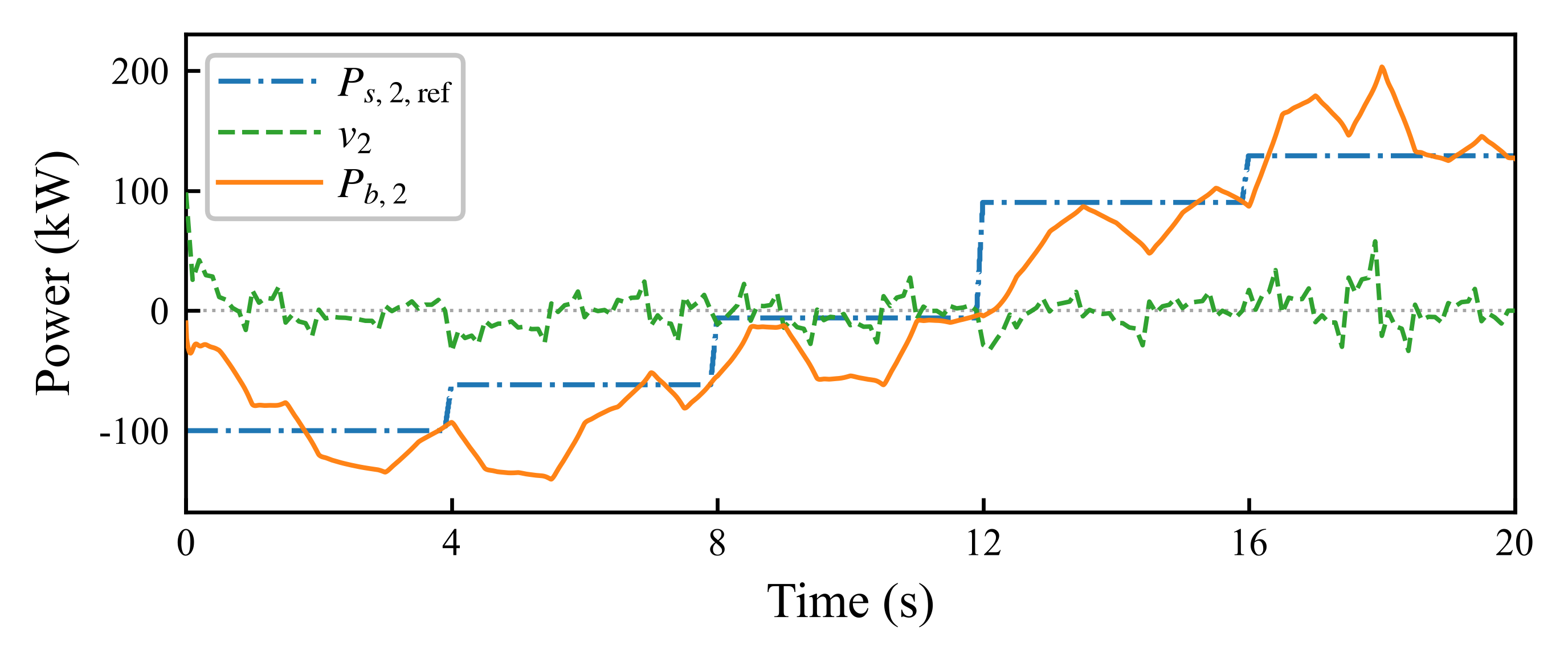}
		\caption{SFR baseline, PFR command, and terminal power of BESS $2$.}
		\label{fig:bess2_power}
	\end{figure}

	To further illustrate the multi-timescale coordination, Figs.~\ref{fig:bess1_power} and Fig.~\ref{fig:bess2_power} present the SFR setpoint $P_{s,i,\mathrm{ref}}$, external PFR command $v_i$, and grid-terminal power $P_{b,i}$ of BESS~1 and BESS~2, respectively. The two units exhibit distinct SFR schedules and fast PFR responses, reflecting the combined effects of their power limits, VSG parameters, and dynamic response characteristics. In particular, the SFR setpoint of BESS~1 reaches its headroom-constrained upper limit of 150~kW in the final period. The upper-level problem schedules SFR setpoints at 4s intervals, while the lower-level problem determines PFR commands at 0.1s resolution. The terminal power $P_{b,i}$ follows the combined command trend while reflecting the second-order BESS dynamics and inherent VSG inertia and damping.  These results illustrate how the bilevel framework coordinates slow-timescale SFR scheduling and fast PFR responses across heterogeneous BESS units, accounting for their operating constraints and dynamic characteristics.
	
%	To further illustrate the multi-timescale coordination, Fig.~\ref{fig:bess1_power} and Fig.~\ref{fig:bess2_power} present the SFR baseline $P_{s,i,\mathrm{ref}}$, external PFR command $v_i$, and grid-terminal power $P_{b,i}$ of BESS~1 and BESS~2, respectively. Owing to their different energy capacities, operating limits, VSG parameters, and dynamic response characteristics, the two BESS units exhibit distinct baseline allocations and
%	fast PFR responses.
%	The upper-level problem updates $P_{s,i,\mathrm{ref}}$ every 4s to allocate slow-timescale regulation power, whereas the lower-level problem updates $v_{i,t}$ every 0.1s to adjust the scheduled baseline. The terminal power $P_{b,i}$ follows the overall command trend while reflecting the second-order BESS dynamics and inherent VSG inertia and damping responses.
	
%	The resulting trajectories confirm that the bilevel framework allocates regulation according to the operating states and dynamic capabilities of the heterogeneous BESS units, thereby coordinating slow-timescale power allocation with fast frequency support.

\section{Conclusion}
\label{sec:conclusion}

{This paper presented a multi-period bilevel optimization framework for coordinating primary and secondary frequency regulation in islanded microgrids via grid-forming BESS. By recasting the bilevel problem into an MPCC using KKT conditions, we developed a barrier-coupled Scholtes--interior-point continuation method with proven C-stationarity guarantees. 
Numerical results on frequency-critical microgrids demonstrated that the proposed method reduces
solution time by 89.3\% and peak frequency deviation by 0.53\%
relative to conventional Scholtes relaxation. 
Future work will focus on multi-area microgrids with high renewable penetration.}

\appendices
\section{Proof of Theorem~\ref{thm:convergence}}
\label{app:convergence}

Since $0<\mu_{\mathrm B}^0<1$, $0<\alpha_\mu<1$, and
$\beta_\mu>1$, \eqref{eq:mu_update} gives
$0<\mu_{\mathrm B}^{r+1}\le
\alpha_\mu\mu_{\mathrm B}^{r}$.
Hence,
$\mu_{\mathrm B}^{r}\le
\alpha_\mu^{r}\mu_{\mathrm B}^{0}\to0$,
and continuity of $\psi$ with $\psi(0)=0$ yields
$\xi^r\to0$.

The relaxed complementarity equality is included in the KKT residual,
and hence
\begin{equation}
\left|
\pi_\ell(\boldsymbol z^r)w_\ell(\boldsymbol z^r)
+\sigma_\ell^r-\xi^r
\right|
\leq\epsilon_r,
\qquad
\ell=1,\ldots,N_c.
\label{eq:relaxed_eq_residual}
\end{equation}
The logarithmic-barrier domain ensures
$\pi_\ell(\boldsymbol z^r)>0$,
$w_\ell(\boldsymbol z^r)>0$, and
$\sigma_\ell^r>0$. Therefore,
\begin{equation}
0
\leq
\pi_\ell(\boldsymbol z^r)w_\ell(\boldsymbol z^r)
\leq
\xi^r+\epsilon_r,
\qquad
\ell=1,\ldots,N_c,
\label{eq:comp_bound}
\end{equation}
and, similarly,
$0<\sigma_\ell^r\leq\xi^r+\epsilon_r$.
Consequently, $\sigma_\ell^r\to0$.

Boundedness ensures that the primal sequence has accumulation points. Fix any such point $\boldsymbol z^\ast$ and select a subsequence converging to it, retaining the index $r$ for simplicity. Taking limits in~\eqref{eq:comp_bound} gives $\pi_\ell(\boldsymbol z^\ast)w_\ell(\boldsymbol z^\ast)=0$. Continuity and positivity along the sequence imply nonnegativity of $\pi_\ell(\boldsymbol z^\ast)$, $w_\ell(\boldsymbol z^\ast)$, and $q_m(\boldsymbol z^\ast)$. The vanishing equality residual further gives $\boldsymbol H(\boldsymbol z^\ast)=\boldsymbol0$, establishing~\eqref{eq:limit_comp}.

For each lower-level convex QP with affine constraints, every optimal primal trajectory admits KKT multipliers, and every primal-dual KKT point has a globally optimal primal trajectory. Thus, every bilevel-feasible trajectory extends to an MPCC-feasible point, and every MPCC-feasible point projects onto a bilevel-feasible trajectory. Since the two formulations have the same feasible primal trajectories and the same upper-level objective, $J_{\mathrm{BL}}^\star=J_{\mathrm{MPCC}}^\star$. Feasibility of $\boldsymbol z^\ast$ gives $J_{\mathrm{MPCC}}^\star\leq J_s(\boldsymbol z^\ast)$, and continuity of $J_s$ yields convergence of the objective values along the selected subsequence. Together, these establish~\eqref{eq:objective_limit}.

To establish C-stationarity, let $\eta_\ell^r$ be the multiplier of the relaxed equality $\pi_\ell w_\ell+\sigma_\ell-\xi^r=0$. Inexact $\sigma_\ell$-stationarity gives
\begin{equation}
	\eta_\ell^r
	=
	\frac{\mu_{\mathrm B}^r}{\sigma_\ell^r}
	+\mathcal{O}(\epsilon_r).
	\label{eq:sigma_stationarity}
\end{equation}
Eliminating $\eta_\ell^r$ gives coefficients $a_\ell^r+\mathcal{O}(\epsilon_r)$ and $b_\ell^r+\mathcal{O}(\epsilon_r)$ for $-\nabla\pi_\ell$ and $-\nabla w_\ell$, respectively, since $\pi_\ell(\boldsymbol z^r)$ and $w_\ell(\boldsymbol z^r)$ remain bounded along the selected
convergent subsequence. The remaining stationarity residual is also $\mathcal{O}(\epsilon_r)$. The assumed multiplier boundedness  permits a further convergent subsequence, along which the residual vanishes and the limiting stationarity equation holds.

From~\eqref{eq:effective_multipliers},
\begin{equation}
a_\ell^r\pi_\ell(\boldsymbol z^r)
=
b_\ell^r w_\ell(\boldsymbol z^r).
\label{eq:effective_multiplier_identity}
\end{equation}
If $\pi_\ell(\boldsymbol z^\ast)>0$, complementarity gives
$w_\ell(\boldsymbol z^\ast)=0$. Boundedness of $\{b_\ell^r\}$ and
\eqref{eq:effective_multiplier_identity} then imply
$a_\ell^\ast=0$. Similarly,
$w_\ell(\boldsymbol z^\ast)>0$ implies $b_\ell^\ast=0$.
The ordinary inequality multipliers $\mu_{\mathrm B}^r/q_m(\boldsymbol z^r)>0$ have multiplier--slack products equal to $\mu_{\mathrm B}^r\to0$, recovering dual feasibility and complementary slackness in the limit. Hence, the limiting system is weakly stationary.

Finally, strict positivity of
$\pi_\ell(\boldsymbol z^r)$,
$w_\ell(\boldsymbol z^r)$, and $\sigma_\ell^r$ gives
\begin{equation}
a_\ell^r b_\ell^r
=
\frac{
(\mu_{\mathrm B}^r)^2
\left[
\sigma_\ell^r
-
\pi_\ell(\boldsymbol z^r)w_\ell(\boldsymbol z^r)
\right]^2
}{
\pi_\ell(\boldsymbol z^r)
w_\ell(\boldsymbol z^r)
(\sigma_\ell^r)^2
}
\geq0.
\label{eq:c_multiplier_product}
\end{equation}
Passing to the limit yields
$a_\ell^\ast b_\ell^\ast\geq0$ for every biactive pair satisfying
$\pi_\ell(\boldsymbol z^\ast)=w_\ell(\boldsymbol z^\ast)=0$.
Thus, $\boldsymbol z^\ast$ is C-stationary for the original MPCC.
%\cite{Scholtes2001,hoheisel2013theoretical,kanzow2015price}.
\hfill\IEEEQED

\bibliographystyle{IEEEtran}
\bibliography{IEEEabrv,ref}

\end{document}